\documentclass[preprint,review,numbers,sort&compress]{elsarticle}
\usepackage[inner=1.0in, outer=1.0in, top=1.0in, bottom=1in, paperheight=11in, paperwidth=8.5in]{geometry}

\usepackage{tikz}
\usetikzlibrary{bayesnet}
\usetikzlibrary{arrows}
\usetikzlibrary{backgrounds}
\usepackage[utf8]{inputenc}
\usepackage{amsfonts}
\usepackage{graphicx}
\usepackage{float}
\usepackage{subcaption}
\graphicspath{{}}
\usepackage{caption}
\usepackage{dirtytalk}
\usepackage{xcolor}
\usepackage{soul}
\usepackage{cancel}

\usepackage{epsfig} 
\usepackage{amsmath} 

\usepackage{algorithm}
\usepackage{algpseudocode}

\title{A probabilistic approach for acoustic emission based monitoring techniques: with application to structural health monitoring}
\author{C.A. Lindley}
\author{M.R. Jones}
\author{T.J. Rogers}
\author{E.J. Cross}
\author{R.S. Dwyer-Joyce}
\author{N. Dervilis}
\author{K. Worden}

\address{Dynamics Research Group, Department of Mechanical Engineering, University of Sheffield, Mappin Street, Sheffield S1 3JD, UK}

\begin{document}

\begin{abstract}
    It has been demonstrated that acoustic-emission (AE), inspection of structures can offer advantages over other types of monitoring techniques in the detection of damage; namely, an increased sensitivity to damage, as well as an ability to localise its source. There are, however, numerous challenges associated with the analysis of AE data. One issue is the high sampling frequencies required to capture AE activity. In just a few seconds, a recording can generate very high volumes of data, of which a significant portion may be of little interest for analysis. Identifying the individual AE events in a recorded time-series is therefore a necessary procedure for reducing the size of the dataset and projecting out the influence of background noise from the signal. In this paper, a state-of-the-art technique is presented that can automatically identify cluster the AE events from a probabilistic perspective. A nonparametric Bayesian approach, based on the \textit{Dirichlet process} (DP), is employed to overcome some of the challenges associated with this task. Additionally, the developed model is applied for damage detection using AE data collected from an experimental setup. Two main sets of AE data are considered in this work: (1) from a journal bearing in operation, and (2) from an Airbus A320 main landing gear subjected to fatigue testing.
\end{abstract}

\maketitle

\section{Introduction}
\label{sec:intro}

A research topic of increasing popularity in \textit{Structural Health Monitoring} (SHM), is the use of \textit{Acoustic Emission} (AE)-based techniques for damage detection. Although AE was first defined by Kaiser in $1950$~\cite{Kaiser1950}, it was only with the development of high sampling-frequency instrumentation and highly-sensitive transducers that the detection of AEs was possible, and an extensive amount of research has since been carried out to better understand the underlying correlation between the physics of AEs and the mechanisms that produce them in mechanical systems~\cite{Hanchi1991,Fan2010,Hase2012,Hellier2013,Feng2019_2,Ma2021}. Additionally, in a more practical context, several promising results have been demonstrated when implementing AE-based monitoring techniques for damage detection in structures and rotating machines. Some examples of this exercise include rolling element bearings~\cite{ChoudhuryA2000Aoae,He2009}, gears~\cite{Feng2019_1}, journal bearings~\cite{Poddar2016,PoddarSurojit2019Dopc}, self-compacting concrete specimens~\cite{Chen2019}, and aerospace structures~\cite{Holford2009,Hensman2011}.

Compared to other forms of monitoring techniques, AE measurements can provide a more sensitive means of detecting the onset of damage, facilitating the earlier diagnosis of an unhealthy structure. This attractive attribute can be of great value for the development of an optimised maintenance schedule. For example, considering an offshore wind turbine, it is more convenient to determine remotely whether maintenance is required, given the associated high cost of having the wind turbine put out of service for a potentially unnecessary repair. Conversely, if the detected damage in the wind turbine suggests that it will fail prior to its next scheduled maintenance, then action can be taken to prevent a potentially-catastrophic outcome. The acquisition of a more sensitive damage-detection system therefore means having more time to plan a suitable course of action.

\begin{figure}[htbp]
    \centering
    \includegraphics[width=0.5\textwidth]{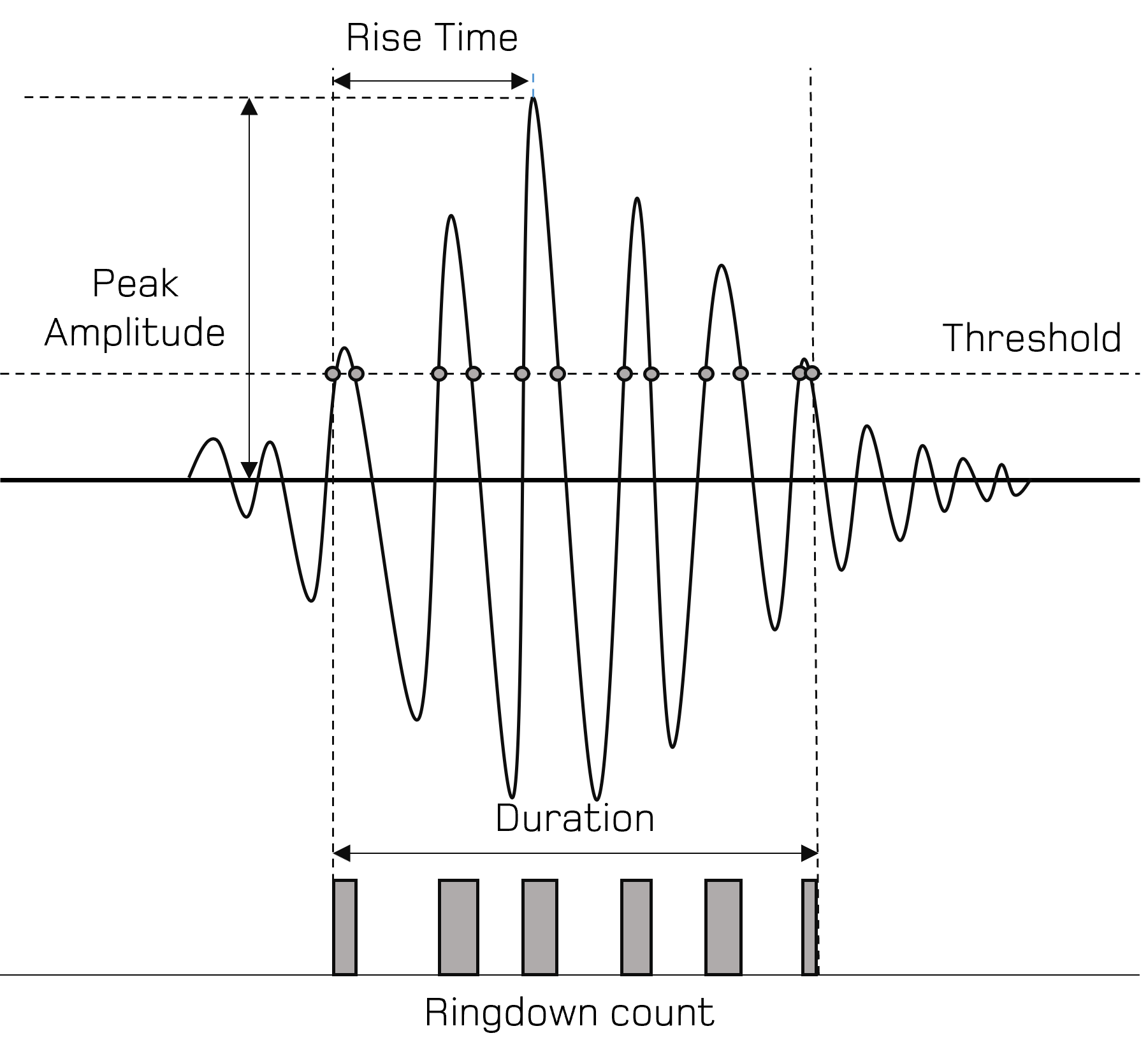}
    \caption{Features of a typical AE. Reproduced from~\cite{Farrar2013}.}
    \label{fig:AE_features}
\end{figure}

Despite its promising advantages in the early detection of damage, there are some added complications associated with AE analysis that need careful attention~\cite{Sikorska2008}. One of these limitations relates to the high sampling frequencies required to record AE signals. In a matter of a few seconds, a recording can generate millions of data-points, making them tedious to store and manipulate. One is therefore likely to rely on feature-extraction methods in order to reduce the overall size of the measured data. The features of interest are those that define the AE, and a few examples of these features can be seen in Figure \ref{fig:AE_features}, for a typical AE waveform.

Various means of feature extraction exist~\cite{Terchi2001}; a common approach is to first identify and isolate the individual AE events/bursts that emerge throughout the recorded signal. The challenge here, however, is to have an algorithm to reliably identify the relevant AE events for analysis, and distinguish them from background noise.

One of the first steps in an AE-based monitoring scheme is to define a threshold from which any voltage crossings are assumed to indicate the presence of some form of AE burst activity. Unfortunately, deciding on the threshold is not as trivial as it may seem. In practice, the threshold may be established from experience, and judging a suitable threshold value will depend on the properties and geometrical constraints of the propagating medium. This approach is limited in that it is specific to the application at hand; the operator must carefully select an optimal threshold to capture the AE events of interest. 

Nevertheless, this exercise is not only meaningful for reducing the overall size of the dataset, but also for retaining AE events carrying information of the health-state of the structure. The underlying features that characterise the isolated AE events are somewhat influenced by their source, and some correlation will exist based on the type of damage or physical mechanism that generates them.

For example, several attempts have been pursued to correlate AE features - particularly the \textit{ringdown count} (Figure \ref{fig:AE_features}) - with fracture-mechanical parameters, such as the stress-intensity factor of structures subjected to cyclic loading~\cite{Morton1973,Bassim1994,Berkovits1995,Rabiei2014}. These studies are based on Paris-Erdogan-type laws~\cite{Paris1963}, to make estimates of the remaining fatigue life of structures, and demonstrate that count-rates can in fact be used to estimate the growth-rate of cracks. Even if all potential AE events are correctly identified and isolated for analysis, this condition is another difficult challenge commonly encountered in practice, requiring the sources of AE events to be carefully characterised for analysis. This complication is even more of an inconvenience in applications dominated by a rich collection of AE sources and/or excessive amounts of background noise.

In such cases, statistical modelling and pattern recognition techniques can be used to facilitate the analysis of AE data. In~\mbox{\cite{Rippengill2003}}, for example, a pattern recognition approach was adopted to automatically classify AE signals obtained from a box girder of a bridge. The developed methods were able to identify three distinct classes that could distinguish crack-related signals from those generated by frictional processes. Another promising pre-processing technique that builds on this idea can be found in~\mbox{\cite{Sause2012}}, where the proposed method outputs the best combination of AE-features that can provide the greatest separation of classes; therefore, yielding a dataset that is most sensitive to the various sources of AE.

If wished to learn more about the potential sources of AE, localisation techniques can be employed to determine where in the structure high concentrations of AE activity may originate from. This approach has been explored in~\mbox{\cite{Holford2009,Hensman2009,Carpinteri2012}}, where the localisation capabilities of AE-based techniques are exploited to directly characterise AE activity emerging near the developing crack. Despite being a salient aspect of AE-based monitoring, source localisation can be hard to implement, especially in structures with complex geometries~\mbox{\cite{Hensman2010,Jones2021}}.

Overall, it becomes evident that two emerging challenges limit the extent to which AE techniques can generalise across SHM applications: the first, on the correct identification of individual AE events in a recording, and the second, on the analysis and interpretability of AE data. In this paper, a state-of-the-art means of addressing these problems is proposed; a probabilistic framework is developed by treating the generation of AE events as a random process. Adopting this perspective allows one to quantify the probability of finding an AE event in any arbitrary section of a raw time-series. This implementation is fundamentally employed with a \textit{Poisson} distribution, a \textit{Probability Mass Function} (PMF) that will be used here as the building-block for modelling counts extracted from AE signals. Ultimately, a powerful nonparametric Bayesian approach is introduced and implemented, both to find the AE events in the raw signal, and to group them based on their features in an online scheme. This approach is further demonstrated to be applicable for early damage identification in a structure using experimental AE data.

The proposed strategies are described in the following sections, and then demonstrated with a rich AE dataset collected from a journal bearing in operation, and from an Airbus A320 main landing gear subjected to fatigue testing.


\section{A probabilistic perspective for AE-based monitoring techniques}

As highlighted above, determining the threshold for AE events can be an assiduous task that may depend highly on the expertise of the operator. Alternatively, a potentially more robust approach could be employed to have the threshold defined statistically. One may assume the background noise in the signal to be the result of a generating process in which the observations are samples drawn from a Gaussian distribution, and the threshold could then be determined numerically using a Monte Carlo method~\cite{Farrar2013}. Although this approach offers a threshold determined directly from the recorded signal, the resulting selection of AE events may still be heavily corrupted by background noise. For example, if a $99^{th}$-percentile threshold is chosen on the recorded time-series, it means (at the risk of redundancy), having $1\%$ of the data points estimated to lie above that threshold. This deceptively-small percentage could be a problem in scenarios where the sampling frequency is high, such as an AE reading. In the span of a few seconds, millions of data points are recorded, and thus thousands of \say{events} would consequently be taken into consideration for further analysis. This issue means mistakenly identifying a vast number of events deriving from the background noise, when it is likely that only a fraction of them will actually originate from damage.

An attempt to overcome these limitations is explored in this work by introducing a probabilistic framework; the idea is to quantify how certainly a threshold crossing is believed to derive from an AE event. Concretely, after establishing a threshold statistically, and then counting the number of times a signal exceeds the threshold in a given time-frame, one can evaluate the probability of an AE event existing in that time-frame. In the following sections, it will be demonstrated how the Poisson distribution can be conveniently used for this purpose.

\subsection{The Poisson distribution}
\label{subsec:poisson}

The Poisson distribution is a discrete probability distribution used to model the number of events occurring in a given time or space interval. More precisely, it is often used to model counts of rare occurrences, such as radioactive decay or traffic accidents~\mbox{\cite{murphy2013}}. Since the intention is to model the number of threshold-crossings in a given time-frame, and detect an unusually high number of threshold-crossing attributed to an AE-burst, the Poisson distribution arises naturally for this type of problem. The Poisson PMF is defined as,
\begin{equation}
    p(x|\lambda) = \frac{e^{-\lambda}\lambda^{x}}{x!}
    \label{eq:poisson}
\end{equation}
where $x$ is a positive integer that, in this application, corresponds to the counts of threshold-crossings happening within a given time-frame, and $\lambda$ is the rate or expected number of these threshold-crossing happening in the time/space interval.

The value of the parameter $\lambda$ may be unknown, and a Bayesian framework is adopted here to infer the distribution over $\lambda$. The count-rate is therefore modelled according to some prior probability distribution, $p(\lambda)$, that quantifies a prior belief over all possible values of $\lambda$. Bayes' rule can then be applied for inference to recover the posterior $p(\lambda|x)$.

\begin{figure}[htbp]
    \centering
    \begin{subfigure}[b]{0.475\textwidth}
        
        \includegraphics[width=\textwidth]{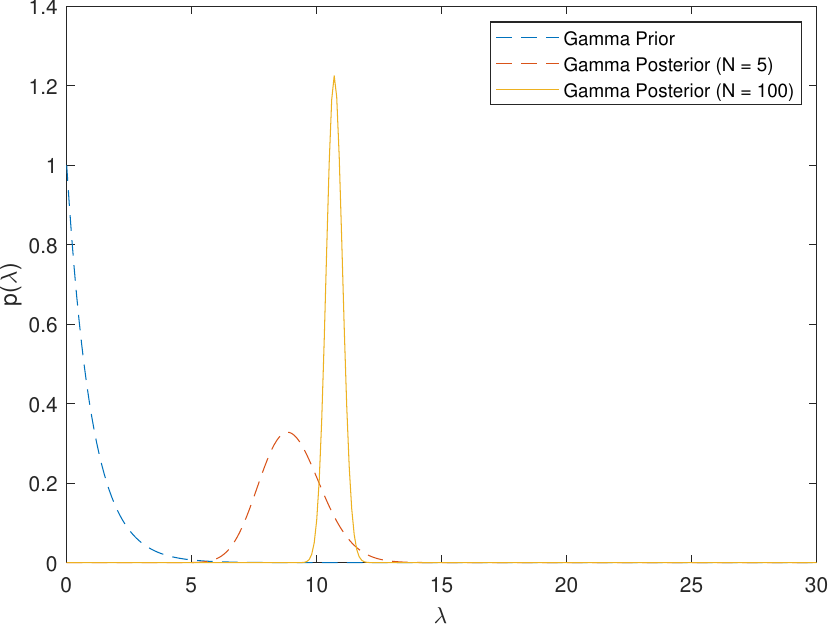} 
        \caption{}
    \end{subfigure}
    \begin{subfigure}[b]{0.475\textwidth}
        
        \includegraphics[width=\textwidth]{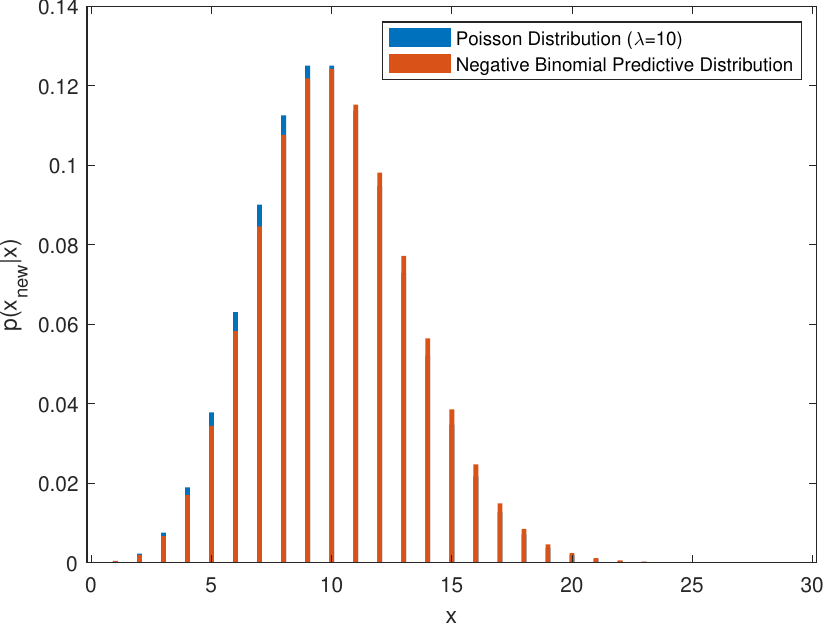} 
        \caption{}
    \end{subfigure}

\caption{(a) Bayesian updating of the Gamma prior $p(\lambda)$, when presented with $N$ samples drawn from a Poisson distribution parameterised by $\lambda=10$. (b) Predictive posterior distribution $p(x_{new}|\mathbf{x})$, evaluated given $N=100$ observations, and compared against the Poisson likelihood $p(\mathbf{x}|\lambda=10)$.}
\label{fig:Bayesian_updates}
\end{figure}

Fortunately, a conjugate prior of the Poisson distribution exists, allowing for a closed-form solution for $p(\lambda|x)$. In order to meet this condition, the chosen conjugate prior must be a Gamma distribution, resulting in another Gamma distribution when combined with the Poisson likelihood~\cite{murphy2013}. The Gamma distribution over $\lambda$ can be defined as,
\begin{equation}
    p(\lambda|a,b) = \frac{b^a}{\Gamma(a)}\lambda^{a-1}e^{-b\lambda}
    \label{eq:gamma_function}
\end{equation}
where $\Gamma(a)$ is the Gamma function that ensures Equation (\ref{eq:gamma_function}) is normalised, and the parameters $a>0$ and $b>0$ correspond to its shape and rate, respectively. These parameters define the functional form of the distribution. In this context, these parameters are intentionally chosen in a flexible manner to align the Gamma distribution with prior beliefs regarding the potential values of $\lambda$. Finally, the predictive distribution for a new observation $x_{new}$, given a some available observations, can then be calculated by taking the expectation of the Poisson likelihood with respect to the posterior distribution,
\begin{equation}
    p(x_{new}|\mathbf{x}) = \mathbb{E}_{p(\lambda|x)}[p(x_{new}|\lambda)] = \int \! p(x_{new}|\lambda)p(\lambda|\mathbf{x}) \, \mathrm{d}\lambda
    \label{eq:predictive_distribution}
\end{equation}
where a set of $N$ observations $\mathbf{x}=\{x_1,\dots,x_N\}$, is considered. Conveniently, equation (\ref{eq:predictive_distribution}) reduces to a negative-binomial distribution. In particular,
\begin{equation}
    p(x_{new}|\mathbf{x}) = \frac{\Gamma(x_n + r)}{x_n!\Gamma(r)}p^r(1-p)^r
    \label{eq:neg_bin}
\end{equation}
where $r = a^*$ and $p = b^*/(b^*+1)$, with the updated parameters $a^*$ and $b^*$ given by,
\begin{align*}
    a^* = N\bar{\mathbf{x}} + a, \quad b^* = \frac{N + b}{N + b + 1}
\end{align*}
where $\bar{\mathbf{x}}$ is the sample mean of the number of available observations $\mathbf{x}$. The probability of observing a certain number of events in a given time interval can be easily calculated from (\ref{eq:neg_bin}), and a meaningful representation of the uncertainty over $\lambda$ is given in the process. An illustration of the Bayesian updating process is shown in Figure \ref{fig:Bayesian_updates}. This process essentially returns some quantified uncertainty on the predictions of new observations, which can be useful for the detection of anomalous data departing from the inferred distribution.

The strategy that follows assumes that the extracted counts from the signal are distributed according to a Poisson distribution. In order to demonstrate how this reasoning can be implemented in practice, the considered AE time-series will be limited to unprocessed recordings in which distinct transient events can be observed occurring sporadically in a continuous stream of background noise. This type of AE signal is often referred to as a \textit{burst signal}~\cite{Terchi2001}, and an example of a typical AE recording of this type can be seen in Figure \ref{fig:AE_led_break}. Under this consideration, it is worth noting that a relatively-higher count-rate can be expected in sections encompassing an AE event.

Now, for a given threshold, one can let the Poisson distribution model the counts extracted from the background noise. The first step would be to construct a dataset $\mathbf{x}$, comprised of uni-dimensional features representing the counts observed in a windowed section that correspond uniquely to background noise, or sections where no obvious AE activity exists. It is then easy to imagine that the counts in each section will be low, and one would hence end-up with an array comprised of a combination of zeros and/or random positive integers near zero. This approach may seem counter-intuitive at first, since the background noise is being modelled rather than the actual AE events, but the results may be easier to interpret when evaluating the \textit{Negative Log-Likelihood} (NLL), of new observations with respect to the already inferred distribution in (\ref{eq:predictive_distribution}),
\begin{equation}
    NLL(x_{new}) = -log(p(x_{new}|\mathbf{x}))
    \label{eq:nll}
\end{equation}
This strategy treats AE events as anomalies in the time-series, and evaluating (\ref{eq:nll}) for new observations should therefore return higher-than-normal values whenever encountering sections in the time-series with AE events.
In order to demonstrate this procedure in practice, a simple lead-break case-study test is now considered.


\subsection{AE event identification in a time-series signal recorded from a lead-break test}
\label{sec:lead_break_test}

\begin{figure}[htbp]
    \centering
    \includegraphics[width=0.9\textwidth]{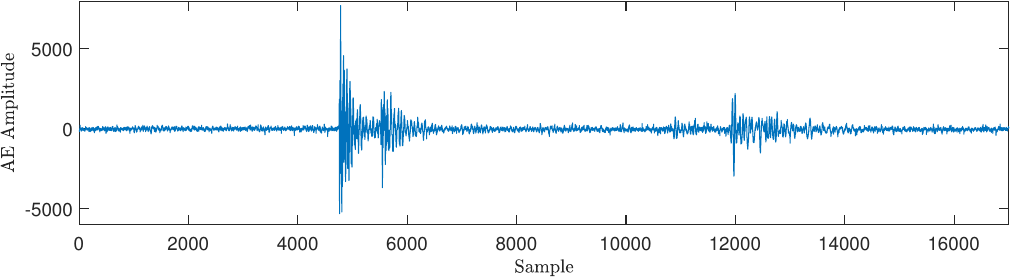}
    \caption{Acoustic emission waveforms from lead break test.}
    \label{fig:AE_led_break}
\end{figure}

A common approach to simulate convincing AE events is by breaking pencil-leads on an elastic surface. This type of source was simulated here with a Hsu-Nielson device~\cite{hsu1977}, and then recorded with a 40dB WB Mistras AE sensor attached to the same surface. Figure \ref{fig:AE_led_break} shows a section of the AE signal recorded during the test. The time-series clearly displays at least two distinct bursts, indicating the instances when the pencil-leads snapped. For the purposes of this demonstration, a $99^{th}$-percentile threshold was employed.

The problem, as highlighted earlier, is that a vast number of \say{events} deriving from the background noise would also be taken into consideration, when only the main events caused by the pencil-lead breaks are of actual interest. Some progress can be achieved by adopting the Bayesian approach described above, where the count-rate $\lambda$, associated with the background noise, can be  assumed to be distributed according to a Gamma distribution, with corresponding parameters set to units ($a = b = 1$), under the assumption that $\lambda$ will be expected to be zero or some other value near zero, given the relatively high threshold established in this demonstration. The reasoning behind this assumption is again based on treating the generated AE-waves as rare events. However, it should be acknowledged that careful tuning of these parameters may be necessary to account for the expected counts given a specific threshold. While the consideration of priors in Bayesian modelling is crucial, exploring the choice of priors for this specific application would require additional extensive work, potentially diverting the reader's attention from the main purpose of this demonstration. Investigating the selection of priors in this context remains a subject for future research.

\begin{figure}[htbp]
    \centering
    \begin{subfigure}[b]{0.9\textwidth}
        \centering
        
        \includegraphics[width=\textwidth]{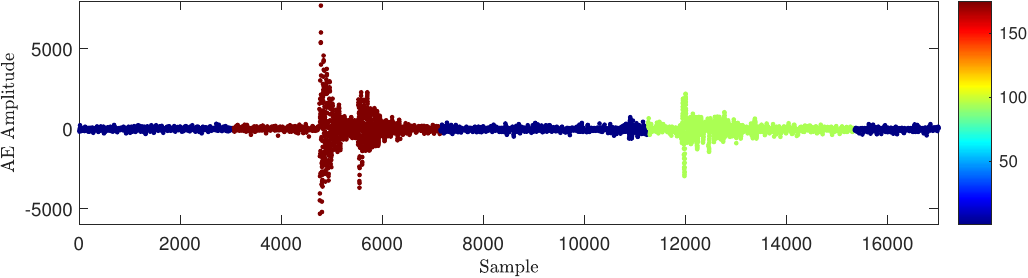} 
        \caption{}
        \label{fig:nwind_4650}
    \end{subfigure}
    \begin{subfigure}[b]{0.9\textwidth}
        \centering
        
        \includegraphics[width=\textwidth]{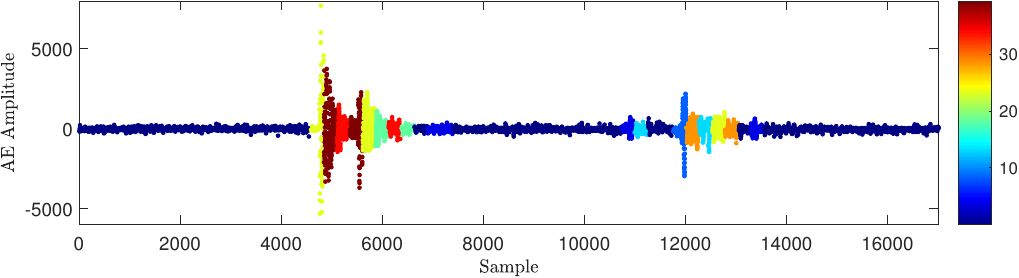} 
        \caption{}
        \label{fig:nwind_256}
    \end{subfigure}

\caption{Negative log-likelihood results on lead-break dataset. The colour bar represents the likelihood of an event existing in a given time interval. Window lengths were (a) $n=4096$ and (b) $n=256$ data points. The threshold used corresponds to the $99^{th}$ percentile of the signal.}
\label{fig:poisson_led_break}
\end{figure}

After windowing the entire time-series with a non-overlapping square function, 20 samples were extracted and used to infer the posterior distribution. It is reiterated here that these samples are counts from sections known to lack any form of meaningful AE activity. The probability of observing $x$ number of threshold crossings in a given section was calculated using equation (\ref{eq:neg_bin}). In order to quantify the confidence of an AE event existing within a section, equation (\ref{eq:nll}) was evaluated on each of the remaining sections. The results can be seen in Figure \ref{fig:poisson_led_break}, where colour-bars representing the NLL were included. By looking at Figure \ref{fig:nwind_4650}, the sudden increases in the NLL can be interpreted as an anomaly being detected, which in turn happen to correspond to the AE events of interest.

An important point worth discussing is the immediate limitation that becomes evident when assessing the sensitivity of the model to the window-length. The results presented in Figure \ref{fig:nwind_4650} seem more promising because a window-length comprised of $4096$ sample-points was chosen beforehand to match the length of the main AE event. Unfortunately, knowing in advance the length of all potential AE events is unlikely to be case, specially when dealing with more complex applications. The effectiveness of this method clearly depends on the length of the sliding window, as demonstrated in Figure \ref{fig:nwind_256}, where a window length of $256$ sample-points was chosen. By having a window-length shorter than the duration of the AE event, the model interprets several components in what should be a single event. The upside of establishing a shorter window-length is that the onset of the events are captured more precisely. It is clear that a validating step would become essential to determine the optimal window-length. In both cases, nonetheless, the background is mostly identified as noise.

Before concluding this section, it may be important to acknowledge that this exercise has been conducted under the assumption that the noise is purely acoustic. However, in practice, there may be other sources of background noise, such as AE activity induced by \textit{Electronic Magnetic Interference} (EMI)~\mbox{\cite{Scruby1987}}, that could introduce complexities to the analysis, necessitating more elaborate monitoring strategies. Exploring this issue further is beyond the scope of this paper, but it certainly merits further investigation.



\section{Nonparametric clustering approach for AE event identification and feature extraction}
\label{sec:nonparam_approach}

The lead-break demonstration provides some insights into the practical implementation of a Poisson distribution for modelling AE data. An interesting development to this approach can be made by extending the model into a mixture of Poisson distributions. The idea is to have the events of interest modelled by a set of independent distributions, each characterised by a different mechanism responsible for the generation of the observable AE events. In doing so, the approach now aims to infer the parameters of the various components comprising the mixture model.

This idea inevitably leads to the problem of identifying AE events in \textit{burst-type} signals, requiring the determination of a suitable window-length (Section \ref{sec:lead_break_test}). Before dwelling on ways in which the robustness of the model can be improved, some background on mixture models will be covered so that this paper is as self-contained as possible. The interested readers can refer to~\cite{murphy2013} or~\cite{Bishop2006} if wishing to learn more on the machine-learning methods covered in this work.


\subsection{Finite Poisson Mixture Model}

Recalling the pre-processing steps followed so far, a feature vector $X$ can be constructed by extracting the number of threshold crossings observed in subsections of a raw AE time-series, encapsulated by a sliding step-window of length $n$. As the window slides through the signal, the number of threshold crossings should vary with changes in the AE activity. The distribution of $X$ will be such that a single Poisson distribution may poorly represent the observations. Therefore, a sensible choice may be to instead model the observations as draws from a combination of several independent Poisson distributions.

First, a multinomial distribution is proposed over a vector of mixing proportions $\boldsymbol{\pi}$, in which each element $\pi_k$, is a mixing coefficient defining the probability that an observation belongs to the $k^{th}$ group. The total number of groups $K$, is predefined, and $\sum_K \pi_k = 1$. It should be noted that the collection of data-points assigned to a distribution in the mixture model will be referred to as a \say{group}. In the literature, the term group is also used interchangeably with \say{cluster}.  

Part of the inference procedure is to derive a latent variable $\textbf{z}$, representing the assignment of a data-point to one of the groups in the mixture model (i.e.\ its labels).
If each group is defined by a unique Poisson distribution with count-rate $\lambda_k$, then each observation $x_n$ can be modelled in the following form,
\begin{equation}
    \begin{split}
        z_n &\sim \mbox{Mult}(\boldsymbol{\pi}) \\
        x_n|z_n &\sim \mbox{Poi}(\lambda_{k})
    \end{split}
\end{equation}
All the model parameters can be then determined efficiently via \textit{Expectation-Maximisation}~\cite{Bishop2006}, resulting in a maximum-likelihood solution, given the data that have been observed so far. Although this model may be a better representation of the observations, it can still be limited in practice. If only a few data points were available, having the local parameters estimated empirically could result in the model not generalising well in the presence of new observations. 

Another consideration is in deciding autonomously on the number of groups the mixture model should have. A possible solution is to assume that the observations can be represented by two unique groups; that is, one characterised by the relatively low count-rate corresponding to background noise, and another that accounts for all the possible higher count-rates that may correspond to AE events. The sequence of sections in the time-series should then be grouped into one in which some form of AE activity exists, and another in which minimal or no events are found. By discarding the latter, one is left with a series of AE events ready to be processed for feature extraction, as required by the application at hand.

However, it might be more convenient to take a step further and split the detected AE events into several groups, in order to get a better insight into their respective sources. For example, it might be better to know whether an AE event corresponds to a specific damaged component, rather than plainly categorising it together with all the rest. The problem, however, is on deciding \textit{a-priori} the number of groups, since one would need to consider every possible condition that a structure or machine might encounter during its lifespan. One way to circumvent this problem is by having the model autonomously create new groups whenever it may be deemed necessary. This approach will require yet another modification, to provide the mixture model the flexibility to account for a possibly \textit{infinite} number of groups.


\subsection{Dirichlet Process Poisson Mixture Model}

\begin{figure}
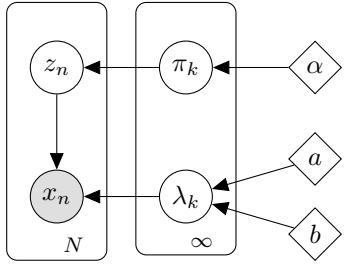

    \centering
    \tikz{
   \node[obs] (x) {$x_n$};%
   \node[latent,right=of x,fill] (y) {$\lambda_k$}; %
   \node[latent,above=of x] (z) {$z_n$}; %
   \node[det,right=of y,yshift=.5cm] (a) {$a$}; %
   \node[det,right=of y,yshift=-.5cm] (b) {$b$}; %
   \node[latent,right=of z] (pi) {$\pi_k$}; %
   \node[det,right=of pi] (alpha) {$\alpha$}; %
   \plate [inner sep=.30cm, yshift=.2cm] {plate1} {(x)(z)} {$N$}; %
   \plate [inner sep=.30cm, yshift=.2cm] {plate2} {(y)(pi)} {$\infty$}; %
   \edge {y,z} {x}  
   \edge {a} {y}
   \edge {b} {y}
   \edge {pi} {z}
   \edge {alpha} {pi}
   }
   \caption{Graphical model of the infinite Poisson mixture model.}
   \label{fig:DPPMM_graphical_model}
\end{figure}

Dirichlet Processes (DPs)~\cite{Ferguson1973}, are powerful stochastic processes with the ability to assign random probability measures to a finite partition $\{A_1,...,A_m\}$ of a continuous measurable space $\Theta$. A DP can be defined by letting $\Theta$ be distributed according to some \textit{base} distribution, $G_o$. A random measure over $G_o$ can be represented by the vector $G=\{G(A_1),...,G(A_m)\}$, which can be said to be drawn from a DP if $G$ is distributed as a finite-dimensional Dirichlet distribution with parameters $\{\alpha G_o(A_1),...,\alpha G_o(A_m)\}$. That is, $G \sim DP(\alpha,G_o)$ if, 
\begin{equation}
    \{G(A_1),...,G(A_m)\} \sim \mbox{Dir}(\alpha G_o(A_1),...,\alpha G_o(A_m))
\end{equation}
where $\alpha$ is a positive scaling parameter. A powerful application of the DP is when it is used as a prior on the parameters of a mixture model. In particular, the observations $x_n$ are assumed to arise from the following generative process,
\begin{align*}
    G|{\alpha,G_o} &\sim \mbox{DP}(\alpha,G_o)  \\
    \theta_n|G &\sim G                   \\
    x_n|\theta_n &\sim p(x_n|\theta_n)
\end{align*}
where $\theta_n$ can now be interpreted as the parameters of the components in the mixture model. A graphical representation of the infinite Poisson mixture model is shown in Figure \ref{fig:DPPMM_graphical_model}. By establishing a DP-prior on the mixture model, a distribution over $K$ can also be inferred from the observations directly as part of the learning procedure, making DPs an attractive solution in SHM/CM when faced with a dataset comprised of a collection of unknown conditions. A detailed introduction to the infinite mixture model can be found in~\cite{Rasmussen1999}, and examples of its use in SHM can be found in~\cite{Rogers2019,Wickramarachchi2022}.

To begin, it will be necessary to make the inference over the parameters Bayesian. This step will lead nicely into the incorporation of a DP-prior over $K$. Firstly, the priors over $\lambda_k$ are considered. As already reviewed in Section \ref{subsec:poisson}, the Gamma distribution is chosen once again to represent the prior belief on $\lambda$. Assuming that the rate values, $\lambda_k$, of each group are independent, allows the specification of the joint density,
\begin{equation}
    p(\lambda_1,...,\lambda_K|a,b) = \prod_{k=1}^K p(\lambda_k|a,b)
\end{equation}
Now, the vector $\boldsymbol{z}_n$ is characterised according to a multinomial distribution parameterised by $\boldsymbol{\pi}$,
\begin{equation}
    p(\boldsymbol{z}_n|\boldsymbol{\pi}) = \prod_k \pi_k^{z_{nk}}
\end{equation}
where the normalising constant simplifies to unity, and the expression is reduced to the probability of assigning a group. A Dirichlet distribution is chosen as a suitable prior on $\pi$, and is also conjugate to the multinomial distribution. The probability density function of the Dirichlet distribution is given by,
\begin{equation}
    p(\boldsymbol{\pi}|\boldsymbol{\alpha}) = \frac{\Gamma(\sum_k \alpha_{k})!}{\prod_k \Gamma(\alpha_k)} \prod_k \pi_k^{\alpha_{k}}
    \label{eq:dir_dist}
\end{equation}
which is parameterised by $\boldsymbol{\alpha}=\{\alpha_1,...,\alpha_k\}$. Having defined the densities on the parameters, the joint posterior likelihood is given by,
\begin{equation}
p(\boldsymbol{Z},\boldsymbol{\pi},\lambda_1,\cdots,\lambda_k|X,\boldsymbol{\alpha},a,b) \propto \left[ \prod_n\prod_k (\pi_k p(x_n|\lambda_k))^{z_{nk}} \right] \left[\prod_k p(\lambda_k|a,b) \right]p(\boldsymbol{\pi}|\boldsymbol{\alpha})
\label{eq:joint_post}
\end{equation}
The conditional distributions for each parameter are proportional to the joint distribution, and can be derived analytically to give a Gibbs sampler~\cite{Bishop2006}, to solve equation (\ref{eq:joint_post}). However, it will be worth marginalising out $\boldsymbol{\pi}$ and $\boldsymbol{\lambda}$ from the joint posterior to instead implement a collapsed Gibbs sampler~\cite{MacEachern1994}. This last step will not only improve the robustness of the sampler, but it will be a necessary one to have the model extend to an infinite number of components. The collapsed distribution is finally expressed as,
\begin{equation}
p(z_{nk}=1|\boldsymbol{Z}^{-n},X^{-n},\boldsymbol{\alpha},a,b) \propto \frac{c_k^{-n} + \alpha_k}{\sum_j^K c_j^{-n} + \alpha_j}p(x_n|\boldsymbol{Z}^{-n},X^{-n},a,b)
\label{eq:collapsed_post}
\end{equation}
where,
\begin{equation}
    p(x_n|\boldsymbol{Z}^{-n},X^{-n},a,b) = \mbox{NB}(\boldsymbol{r},\boldsymbol{p})
    \label{eq:neg_bin_non_empty}
\end{equation}
and,
\begin{align*}
\boldsymbol{r} = a + \sum_{m \ne n}z_{mk}x_m, \quad \boldsymbol{p} = \frac{\sum_{m \ne n}z_{mk} + b}{\sum_{m \ne n}z_{mk} + b + 1}
\end{align*}
The new notation $c_k^{-n}$ corresponds to the count of all but the $n^{th}$ element in the $k^{th}$ group, $\boldsymbol{Z}^{-n}$ to the set of all assignments except for the $n^{th}$ one, and $X^{-n}$ to the set of all the other observations; the superscript $-n$ does not represent a power here. The expression in (\ref{eq:collapsed_post}) is also simplified by considering the value where the $k^{th}$ element of $\boldsymbol{z}_n$ is equal to unity, while all others are zero. A full derivation is provided in \ref{appex:derivation_of_p_z}. In this form, $\boldsymbol{z}_n$ can now be resampled directly from equation (\ref{eq:collapsed_post}). The two parts of this expression can be interpreted as comprising a new prior,
\begin{equation}
    p(z_{nk}=1|\boldsymbol{\alpha},\boldsymbol{Z}^{-n}) = \frac{c_k^{-n} + \alpha_k}{\sum_{j = 1}^K c_j^{-n} + \alpha_j }
\end{equation}
and a likelihood defined by the negative binomial density with parameters $\boldsymbol{r}$ and $\boldsymbol{p}$. In this form, an infinite number of components can be managed, if the parameters $\alpha_k$ in (\ref{eq:dir_dist}) are all set to the same value. The Dirichlet prior can then be expressed as,
\begin{equation}
    p(\boldsymbol{\pi}|\alpha) = \mbox{Dir}(\alpha/K,...,\alpha/K)
\end{equation}
and the sampling prior for $z_{nk}=1$ becomes,
\begin{equation}
    p(z_{nk}=1|\alpha,\boldsymbol{Z}^{-n}) = \frac{c_k^{-n} + \alpha/K}{\alpha + N - 1}
    \label{new_prior_2}
\end{equation}
where the fact that $\sum_{j=1}^K\alpha/K=\alpha$ and $\sum_{j=1}^Kc_j^{-n}=N-1$ were used. Now, by letting $K\to\infty$, equation (\ref{new_prior_2}) reduces to,
\begin{equation}
    p(z_{nk}=1|\alpha,\boldsymbol{Z}^{-n}) = \frac{c_k^{-n}}{\alpha + N - 1}
\end{equation}
This new expression defines the prior probability of the $n^{th}$ data point going into the $k^{th}$ group, and it is proportional to the number of members that currently exist in that group. So far, the new prior only accounts for groups in which members exist, but in this framework, there is always a non-zero probability for a data-point to be assigned to a new \say{empty} component that is yet to be created. The probability of a new member finding its place in one of these empty groups can be easily derived by computing $1$ minus the total probability of it going into any of the non-empty components. Hence,
\begin{equation}
    p(z_{nk_*}=1|\alpha,\boldsymbol{Z}^{-n})=1-\sum_{k=1}^{K} \frac{c_{k}^{-n}}{\alpha + N - 1} = \frac{\alpha}{\alpha + N - 1}
\end{equation}
where the subscript $*$ in $k_{*}$ denotes an empty component. One can notice that the prior defines a probability of a new member either going to a non-empty component proportional to the number of its members, or to one of an infinite number of empty components proportional to $\alpha$. Implementing this reasoning may begin by assigning all observations to the same component, and then resampling the assignments of each observation with probability,
\begin{equation}
   p(z_{nk}=1|\alpha,\boldsymbol{Z}^{-n})=
   \begin{cases}
    \frac{c_{k}^{-n}}{\alpha + N - 1}    & \text{for $c_k^{-n}>0$}\\
    \frac{\alpha}{\alpha + N - 1}       & \text{for $c_k^{-n}=0$}
   \end{cases}
   \label{new_prior_3}
\end{equation}
After incorporating the data, an observation is assigned by sampling from $p(z_{nk}=1|...)$. 

When accounting for an infinite number of components $(K\to\infty)$, one can move from having a prior that determines how the observations group together among a fixed number of components, to one that can partition the data in any possible number of them. It should be noted at this point that the term \say{component} has been introduced to refer to the collection of all groups that are \say{empty} and \say{non-empty}. The former corresponds to groups that are yet to have data-points assigned to them, and the latter to those that already exist in the model. 

Finally, the result yields an infinite Gibbs sampling procedure that involves resampling each $\boldsymbol{z}_n$ according to the probabilities,
\begin{equation}
   p(z_{nk}=1|...)\propto
   \begin{cases}
    c_{k}^{-n}p(x_n|\boldsymbol{Z}^{-n},X^{-n},a,b)  & \text{for non-empty components}\\
    \alpha p(x_n|a,b)                   & \text{empty components}
   \end{cases}
   \label{eq:p_z_nk}
\end{equation}
where $p(x_n|\boldsymbol{Z}^{-n},X^{-n},a,b)$ is given by equation (\ref{eq:neg_bin_non_empty}), and $p(x_n|a,b)$ by the expectation of the likelihood, $p(x_n|\boldsymbol{\lambda})$, with respect to the prior $p(\boldsymbol{\lambda}|a,b)$,
\begin{equation}
    p(x_n|a,b) = \int \! p(x_n|\boldsymbol{\lambda})p(\boldsymbol{\lambda}|a,b)\, \mathrm{d}\boldsymbol{\lambda}
    \end{equation}
which, in this instance, is possible to evaluate because $p(x_n|\boldsymbol{\lambda})$ is a Poisson distribution parameterised by $\boldsymbol{\lambda}$, and $p(\boldsymbol{\lambda}|a,b)$ is its conjugate Gamma prior.

The resampling procedure can now be dictated as follows:
\begin{enumerate}
\item All elements are assigned to a non-empty component $(K = 1)$.
\item One element is removed from its component and its likelihood evaluated with equation (\ref{eq:p_z_nk}). The component is eliminated if the element was unique in that component.
\item If the unassigned element is more likely to belong to an empty component, then a new component is created, and the data point is assigned to that component.
\item One sweep is completed when each data point has been evaluated. The whole process is then repeated from Step Two, until the posterior converges to a solution.
\end{enumerate}
It must be noted that inferring (\ref{eq:p_z_nk}) is not limited to a collapsed Gibbs sampler, and variational methods have been developed for inference of Dirichlet process mixture models~\cite{Blei2006}. The sampling method here was preferred for its simpler intuition. The DP-PMM will be now demonstrated to model the AE events in a time-series signal.



\section{AE event identification in time-series signals}
\label{sec:AE_detection_strategy}

Recovering AE events from a raw time-series can be a challenging task, particularly as the behavior of an event will differ from one to the next, and also because consecutive events may also overlap. As mentioned earlier, identifying and extracting these events may be an assiduous task, but a meaningful one, as it should reveal information on the health-state of the machine being monitored.

By introducing a DP prior, the mixture model can now adapt to the complex nature of the AE analysis. Compared to the methods already described, an infinite mixture model may be better suited for the detection of individual AE events in the time-series. Even without knowing the exact nature of the generated AEs, this approach should assign the various events into groups based on their features. Given that the feature of interest here is the count of an event, these groups are therefore modelled by a set of independent Poisson distributions.

\begin{figure}[htbp]
    \centering
    \includegraphics[width=0.95\textwidth]{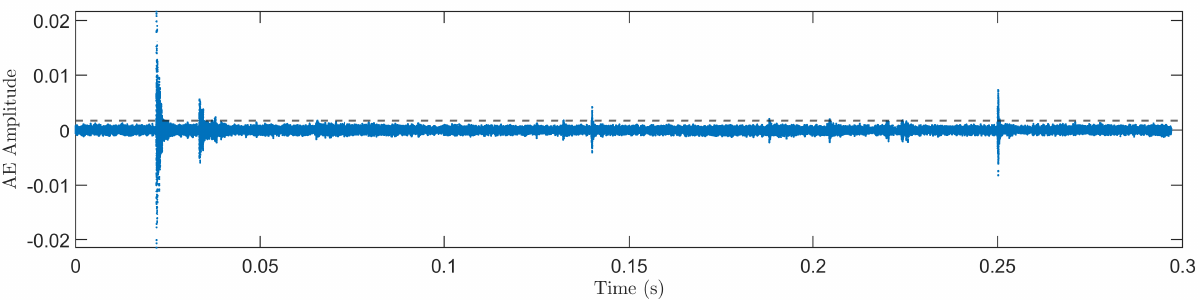}
    
    \caption{AE readings from a plain journal bearing operating under hydrodynamic lubrication at a rotational speed of $150rpm$ and applied static load of $5kN$. The dashed line corresponds to a $99.5^{th}$ percentile threshold.}
    \label{fig:AE_150_section}
\end{figure}

The AE waveform shown in Figure \ref{fig:AE_150_section} will be considered for this demonstration. This waveform corresponds to an AE time-series recorded from a journal bearing operating at $150rpm$ and under constant axial load of $5kN$. For feature extraction, a sliding window approach is again considered. As demonstrated in Section \ref{sec:lead_break_test}, the immediate challenge of this approach is finding a suitable window-length; if too long, it will be less likely to discern individual events and their counts. Conversely, if too short, a single event could mistakenly be interpreted as a collection of them. Under these considerations, the window-length should at least match the duration of the shortest event in the signal. However, it may be impossible to know this information beforehand, and some form of validation would be needed to find the best parameters. If the geometry and materials of the propagating medium are simple enough, some insight can be gained by simulating the potential events. Since this information was not readily available, a step function with a length of $n=2048$ sample points $(\Delta t = 0.002048s)$ was chosen for the windowing procedure. The choice of this specific length was based on~\mbox{\cite{Hensman2011}}, where the same window length was used to record individual AE waves.

Unfortunately, this step alone is limited by the fact that the sliding window is not guaranteed to align with the AE events in the signal. To address this issue, it is necessary to overlap the imposed windows. A window overlap of $87.5\%$ was thus used for this demonstration. The following section dwells on importance of overlapping and demonstrates how it improves the robustness of this approach. Having windowed the signal, the count was then extracted by taking the number of times the signal was found to go over a pre-defined threshold. In this case, a $99.5^{th}$ percentile threshold was defined over the entire waveform. It is worth reiterating that, although a hard-threshold is established here, the probabilistic approach should alleviate the aforementioned shortcomings of having a statistically determined threshold. A threshold crossing will no longer be taken to derive from an AE event; instead, a probability will quantify the (un)certainty of there being an AE event.

\begin{figure}[htbp]
    \centering
    \includegraphics[width=0.66\textwidth]{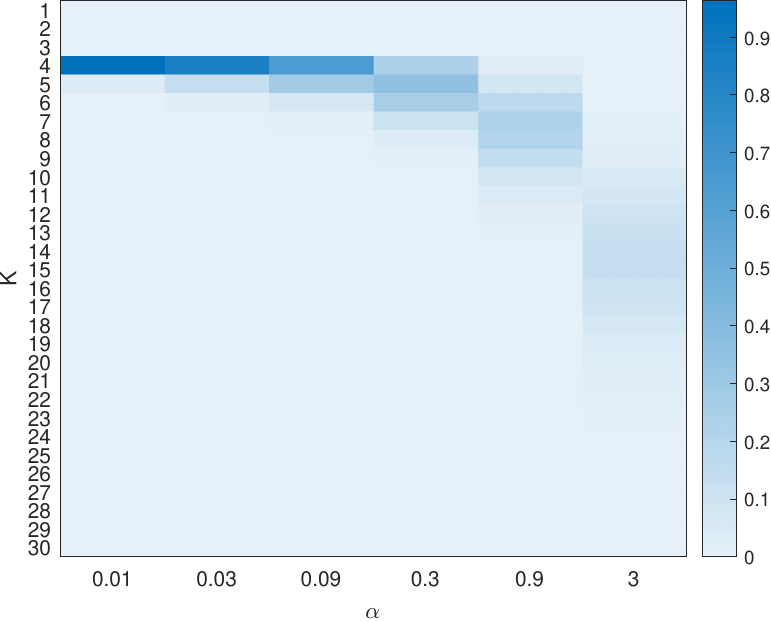}
    
    \caption{Predictive likelihood for the number of clusters $K$ given $\alpha$. The colour-bar indicates the outcome of $p(K|X,\alpha)$.}
    \label{fig:alpha_heatmap}
\end{figure}

With the constructed feature set, the DP-PMM was implemented to infer a suitable partition of the extracted counts. The resulting clusters depend strongly on $\alpha$, and even after having marginalised the underlying parameters in Equation (\ref{eq:p_z_nk}), it is still necessary to find the optimal value for $\alpha$. One way to achieve this is by running the Gibbs sampler multiple times, each time with different values of $\alpha$, and evaluating the probability of data partitioning into $K$ clusters. The probability mass function $p(K|X,\alpha)$, can be computed by keeping track of the number of inferred clusters at each iteration of the Gibbs sampler. The outcomes of this process, given different $\alpha$ values, are presented in Figure (\ref{fig:alpha_heatmap}), highlighting that the most likely partition was attained when $\alpha=0.01$. This finding suggests that the data is optimally represented by four distinct clusters. Consequently, the following section presents the results obtained when $\alpha=0.01$. In all cases, the Gibbs sampler was conducted over $10,000$ iterations.

\subsection{Results and discussion}
\label{sec:results_and_discussion}

\begin{figure}[htbp]
    \centering
    \begin{subfigure}[b]{0.9\textwidth}
        \centering
        \includegraphics[width=\textwidth]{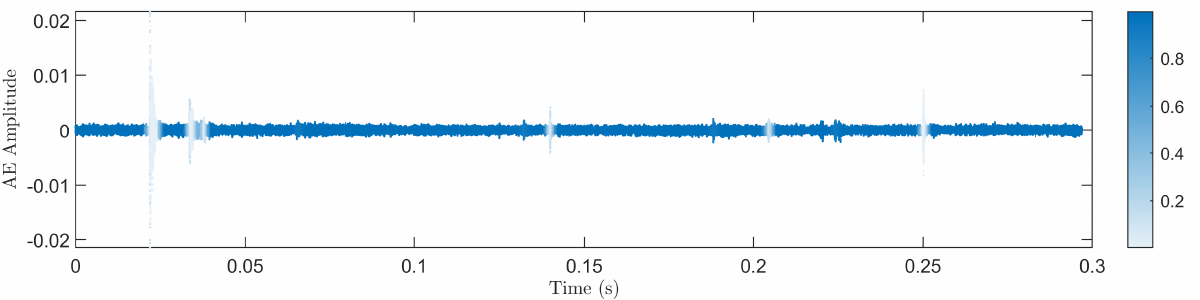}
        \caption{$p(z=1|\dots)$}
    \end{subfigure}
    \begin{subfigure}[b]{0.9\textwidth}
        \centering
        \includegraphics[width=\textwidth]{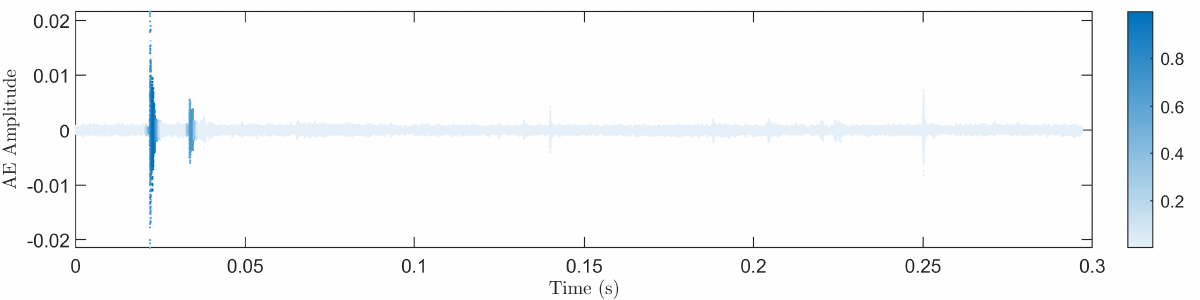}
        \caption{$p(z=2|\dots)$}
        \label{fig:p_znk_2}
    \end{subfigure}
    \begin{subfigure}[b]{0.9\textwidth}
        \centering
        \includegraphics[width=\textwidth]{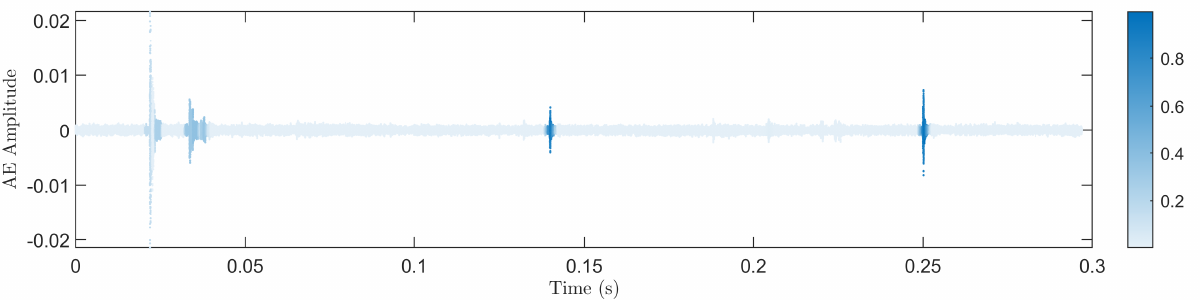}
        \caption{$p(z=3|\dots)$}
    \end{subfigure}
    \begin{subfigure}[b]{0.9\textwidth}
        \centering
        \includegraphics[width=\textwidth]{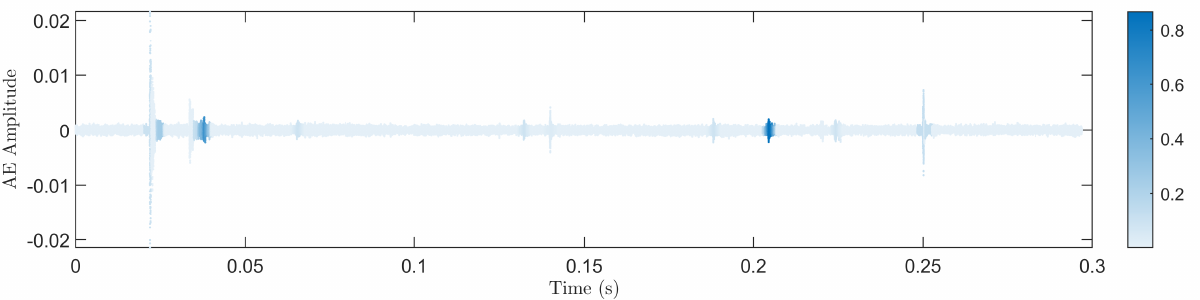}
        \caption{$p(z=4|\dots)$}
    \end{subfigure}
    
    \caption{Visualising the assignment probabilities across the AE signal. The colour-bar corresponds to the predictive posterior probability inferred by the Gibbs sampler.}
    \label{fig:p_z_all}
\end{figure}

\begin{figure}[htbp]
    \centering
    \includegraphics[width=0.9\textwidth]{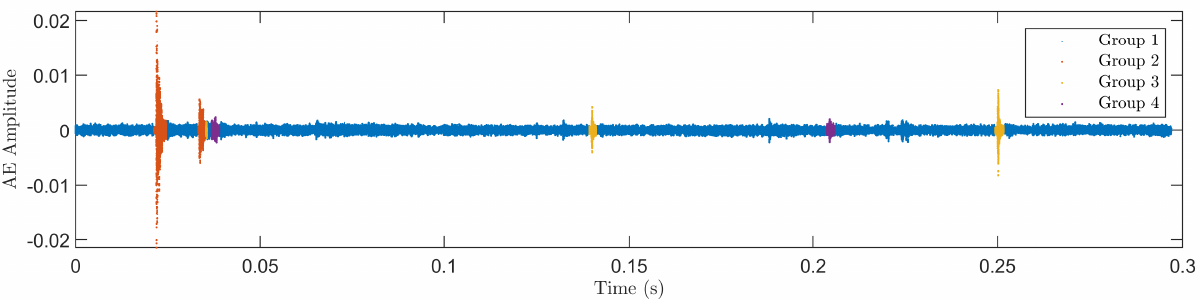}
        
    \caption{Assignment of sections in the signal to the groups that maximise the assignment probabilities. Four distinct groups of AE activity were inferred in this signal.}
    \label{fig:z_max}
\end{figure}

The probability of the assignments inferred by the proposed strategy are shown projected on the signal in Figure \ref{fig:p_z_all}. Visually, it seems as if the main acoustic events were almost undoubtedly detected in all cases. A distinction among the clustered AE events can also be observed in the results. It appears as if the first group mostly captures the background noise, while the remaining groups capture the different AE events found in the signal. If wished to categorise the identified events, one can assign the regions in signal to the group that presents the highest probability. The result of this outcome is illustrated in Figure \ref{fig:z_max}. Not only are these events identified with minimal intervention, but they are also clustered in a way that distinguishes the more \say{imposing} waves, such as the one found in Group $2$, from the \say{smaller} ones, like those found in Groups $3$ and $4$. This result implies that the counts can serve as a representative feature of AE waves.

One of the advantages of having the AE events grouped in this way is that it provides an organised framework from which one can proceed with the analysis. For example, only events from Group $2$ could be considered for extracting their remaining features, such as amplitude, rise-time and energy (among others). This approach significantly reduces the dimensionality of the original dataset and encodes most of the information needed to diagnose the state of the structure or machine being monitored~\cite{Farrar2013}. An emerging defect should affect these features to some extent, giving an anomaly detector an indication that the system has deviated from normal conditions. Alternatively, a new group could develop as damage becomes more prevalent, and one could then query this new component and decide whether it arises from further damage. The nature of the DP-PPM in this context could allow for an online monitoring without having to initiate a new training period.

An additional advantage of this approach is that it provides a principled solution to the problem of deciding on the optimal parameters for feature extraction; most notably, the threshold, window length and amount of overlapping. It has been made evident that these parameters greatly influence the outcome of the model, since these govern the outcome of the feature extraction. Given that the model is Bayesian, it is theoretically possible to also marginalise these parameters. Letting a particularly meaningful combination of these parameters define a model $\mathcal{M}_j$, and having a collection of $J$ finite and discrete models $\mathbf{M}=\{\mathcal{M}_1,\dots,\mathcal{M}_J \}$, the predictive posterior can be evaluated by marginalising over each of the individual models. In particular,
\begin{equation}
    p(\mathbf{z}_{n}|\dots)= \sum_{j = 1}^J p(\mathbf{z}_{n}|\mathcal{M}_j,\dots) p(\mathcal{M}_j)
    \label{eq:marginalisation}
\end{equation}

Assuming an equal contribution from each model, then $p(\mathcal{M}_j) = 1/J$, and the expression in (\ref{eq:marginalisation}) simplifies to the mean of the inferred assignment probabilities obtained from each $\mathcal{M}_j$. There are, indeed, some practical limitations to this proposal. The expense of having to evaluate an exhaustive combination of parameters would be computationally prohibitive. Nevertheless, having to overlap the windows inevitably necessitated the evaluation of Equation (\ref{eq:marginalisation}), and this step was in fact employed to attain the results presented in this section.

\begin{figure}[htbp]
    \centering
    \begin{subfigure}[b]{0.90\textwidth}
        \centering
        \includegraphics[width=\textwidth]{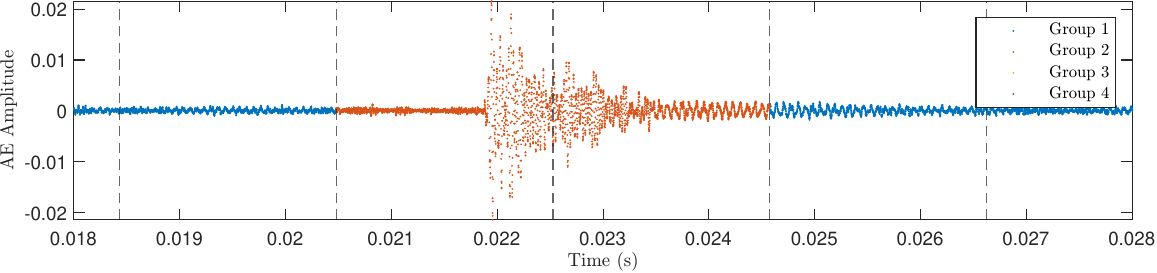}
        \caption{}
        \label{fig:offset1}
    \end{subfigure}
    \begin{subfigure}[b]{0.90\textwidth}
        \centering
        \includegraphics[width=\textwidth]{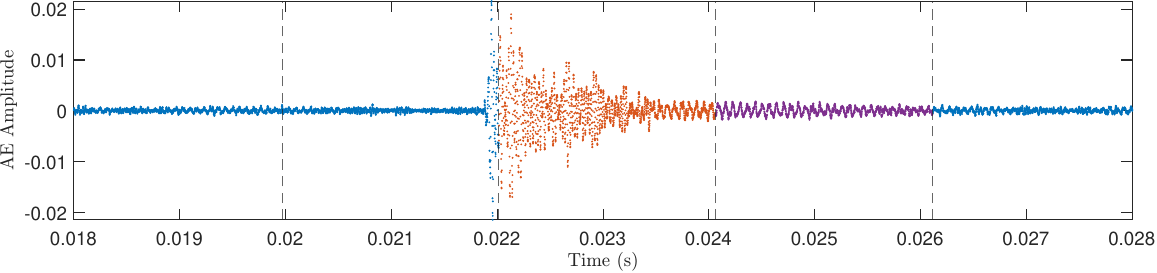}
        \caption{}
        \label{fig:offset2}
    \end{subfigure}
    \begin{subfigure}[b]{0.90\textwidth}
        \centering
        \includegraphics[width=\textwidth]{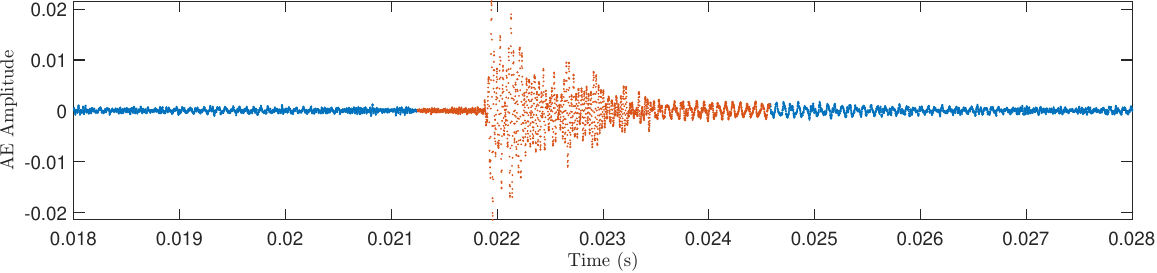}
        \caption{}
        \label{fig:offsetAvg}
    \end{subfigure}
    
    \caption{Close-up view of the main AE event found in Figure \ref{fig:p_znk_2}. The results demonstrate the effects of overlapping the windows for feature extraction, where the assignment of sections in the signal are determined with (a) no overlapping offset, and (b) a $62.5\%$ overlapping offset - $(5/8)N_{\mathrm{window}}$. (c) Resulting assignment of the AE event by averaging over each of the individual models.}
    \label{fig:overlapping}
\end{figure}

An $87.5\%$ overlap (or $12.5\%$ offset) led to the formation of seven distinct sets of feature vectors, with each set considered as a different model realisation $\mathcal{M}_j$. Therefore, this implication required inferring $p(\mathbf{z}_n|\mathcal{M}_j,\dots)$ for each model, before averaging the resulting probabilities. To illustrate this process, Figure \ref{fig:overlapping} shows the effects of the overlapping offsets for the main AE even found in Figure \ref{fig:p_znk_2}, and the result of marginalising $\mathbf{M}$. The top two figures show the assignments inferred given two arbitrary models. It can be seen that somewhat of a good alignment is achieved in the first case, but when an offset is introduced, the partitioning \say{cuts} the event short. Figure \ref{fig:offset2} shows that the section preceding the actual event is assigned to Group $1$. Most of this section is in fact attributed to noise, but an initial portion of the event is captured at the end-edge of the window, misleading the model into assigning this section incorrectly. Although shown otherwise with this particular AE event, offsetting the windows is necessary for all existing events in the signal to be, at least once, somewhat aligned to the windows. Eventually, the evaluation of Equation (\ref{eq:marginalisation}) alleviates the potential discrepancies between models, yielding a balanced assignment and correct identification of the event, as shown in Figure \ref{fig:offsetAvg}. This reasoning can also be extended to incorporate different thresholds and/or window-lengths. However, as mentioned earlier, the inclusion of additional models requires inferring the DP-PMM over a larger dataset, which could be infeasible in practice. This issue will be subject of further investigation in future work, but a brief example of marginalising the threshold can be found in Appendix (\ref{appex:thresh_marginalisation}).

\begin{figure}[htbp]
    \centering
    \includegraphics[width=0.90\textwidth]{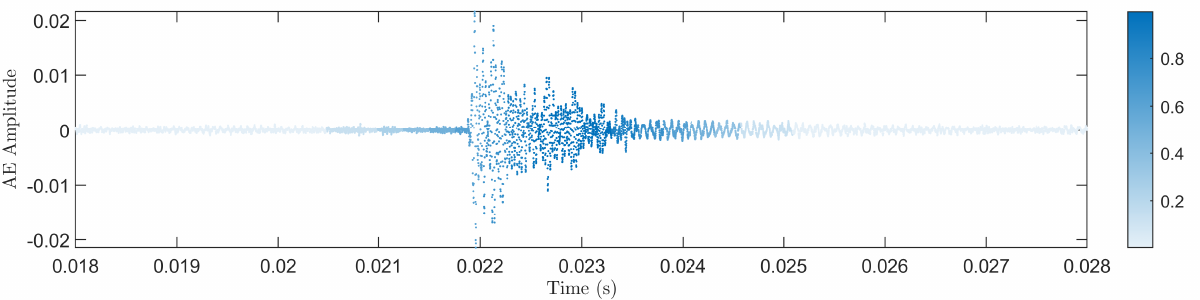}
    \caption{Close-up view of the main AE event found in Figure \ref{fig:p_znk_2} displaying the probability of being assigned to Group $2$.}
    \label{fig:Z2skyStartEnd}
\end{figure}

Finally, an intriguing implication of implementing the probabilistic paradigm for AE identification is that it can naturally indicate the possible start and end points of the AE event by considering the region where the probabilities exceed some minimum value. As shown in Figure \ref{fig:Z2skyStartEnd}, the assignment-probability experiences gradual changes as the AE event unfolds and subsides. This consequence indirectly provides a window-length that potentially adapts to the true duration of the AE event. In comparison to the conventional approach of recording AE waves, this method minimises the risk of losing information about detected waves, as the the recordings of the AE signals are not bounded by fixed lengths.


\section{Application to landing gear}

Identifying AE events in a signal is an important pre-processing step when performing this type of AE-based monitoring. As demonstrated in the previous section, a probabilistic approach may offer several advantages over traditional methods in the identification of AE events, and with the added flexibility of the mixture model, the DP-PMM may be an effective alternative for managing what can be an overwhelming dataset.

An additional application of the DP-PMM in SHM is explored in this section where the modelling approach is employed directly for damage detection, rather than a practical pre-processing step. The experimental study under consideration is the fatigue testing of a $300M$ steel lug welded on an \textit{Airbus} $A320$ main fitting. This study corresponds to one of a series of projects conducted by \textit{The University of Cardiff} in collaboration with \textit{The University of Sheffield}, and under the support of \textit{Messier-Dowty Ltd}. The series of experiments carried out investigated the use of AE techniques for the detection and localisation of fractures in certified landing gears under fatigue testing.

\subsection{Experimental method}

The dataset explored here is an extensive collection of AE waveforms recorded throughout the fatigue test. A loading arm attached to the lug was used to transmit a cyclic load at a rate of $1Hz$ and with a peak amplitude of $5.5kN$. The load was later increased to $6kN$, $6.5kN$ and $7kN$ after $90,000$ cycles, $110,000$ cycles and $138,500$ cycles, respectively. Gradually increasing the load was necessary to promote crack-growth, eventually leading to rupture after $160,000$ cycles ($200,000s$). The test was made more \say{realistic} by also exciting the sliding tube of the main fitting, periodically at a rate of $0.4Hz$. The added contribution of the sliding tube was included to promote the generation of benign AE events, as would be expected outside laboratory conditions. This dataset has been analysed before in~\cite{Hensman2011}, where spatial-scanning statistics were used to detect and localise the source of damage in the landing gear. In the following, the analysis will be limited to the early identification of damage, and how the implementation of nonparametric probabilistic techniques can aid in this task.

The main fitting was mounted with eight \textit{Physical Acoustics Limited (PAL) Nano} $30$ sensors around the cylindrical part of the landing gear (Figure \ref{fig:dm_test_rig}). All sensors were attached to the structure with magnetic-clamps, and brown-grease was used as an acoustic-couplant. Using a $PAL PCI-2$ acquisition system, the measured data were recorded at a sampling rate of $2MHz$. Fixed-size windows of $2048$ samples were recorded upon the signal crossing a pre-established threshold of $43dB$. A memory buffer remained active to include $500$ sample-points before the trigger event of each waveform.


\subsection{Feature selection}

\begin{figure}[htbp]
    \centering
    \begin{subfigure}[b]{\textwidth}
        \centering
        \includegraphics[width=0.7\textwidth]{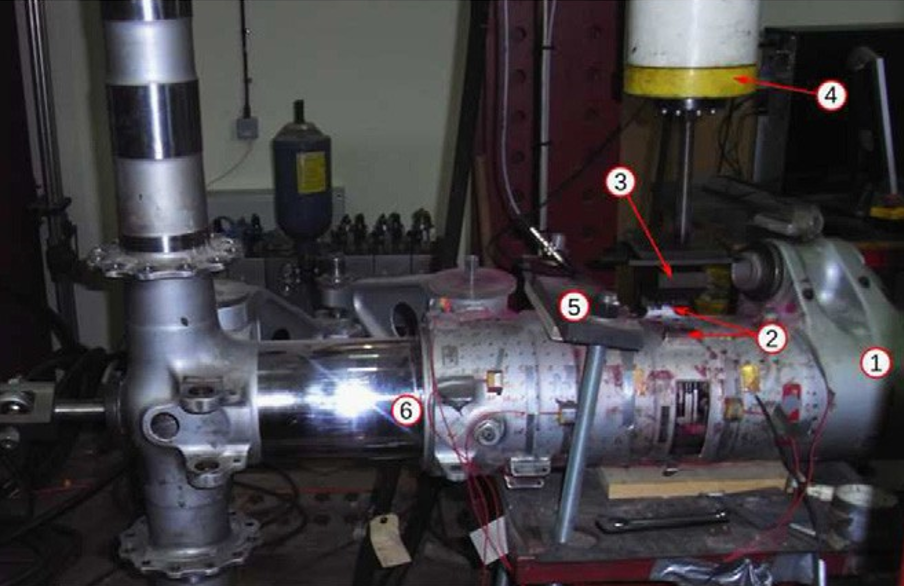} 
        \caption{}
        \label{fig:dm_test_rig}
    \end{subfigure}
    \begin{subfigure}[b]{\textwidth}
        \centering
        \includegraphics[width=0.7\textwidth]{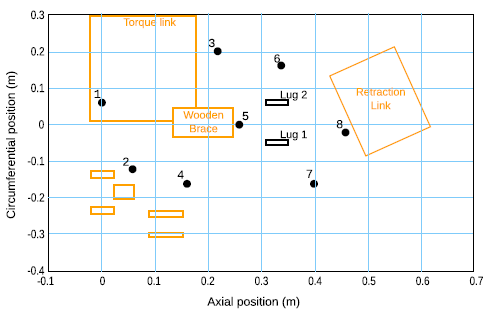} 
        \caption{}
        \label{fig:sensor_arrangement}
    \end{subfigure}

\caption{(a) Experimental set-up and (b) sensor arrangement (red dots) (from~\cite{Hensman2011}). The main landing-gear fitting (1), can be seen to be comprised of the lugs (2), loading arm (3), load actuator (4), restraining wooden brace (5), and a sliding tube (6).}
\label{fig:exp_set_up}
\end{figure}

\begin{figure}[htbp]
    \centering
    \begin{subfigure}[b]{0.9\textwidth}
        \centering
        \includegraphics[width=\textwidth]{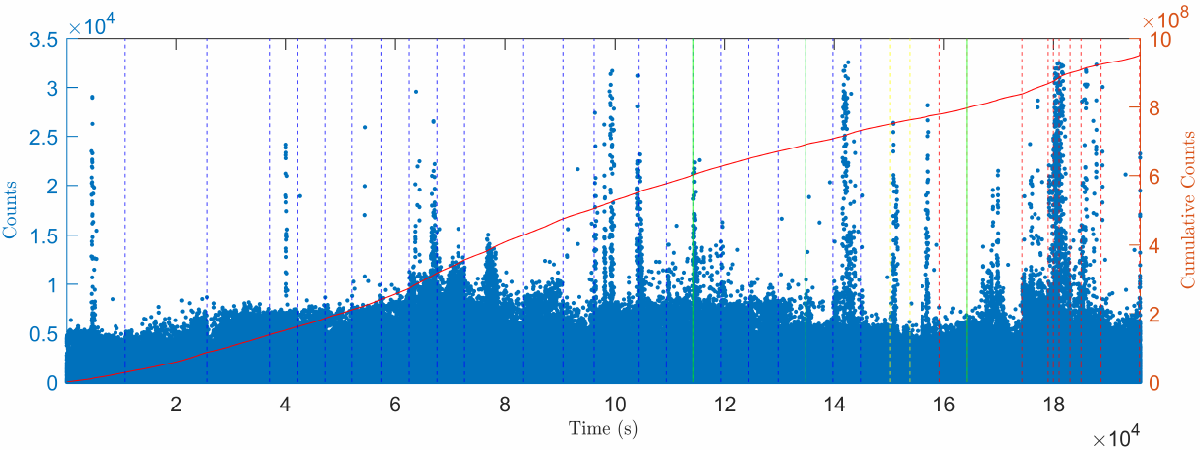} 
        \caption{}
        \label{fig:CompleteSet}
    \end{subfigure}
    \begin{subfigure}[b]{0.9\textwidth}
        \centering
        \includegraphics[width=\textwidth]{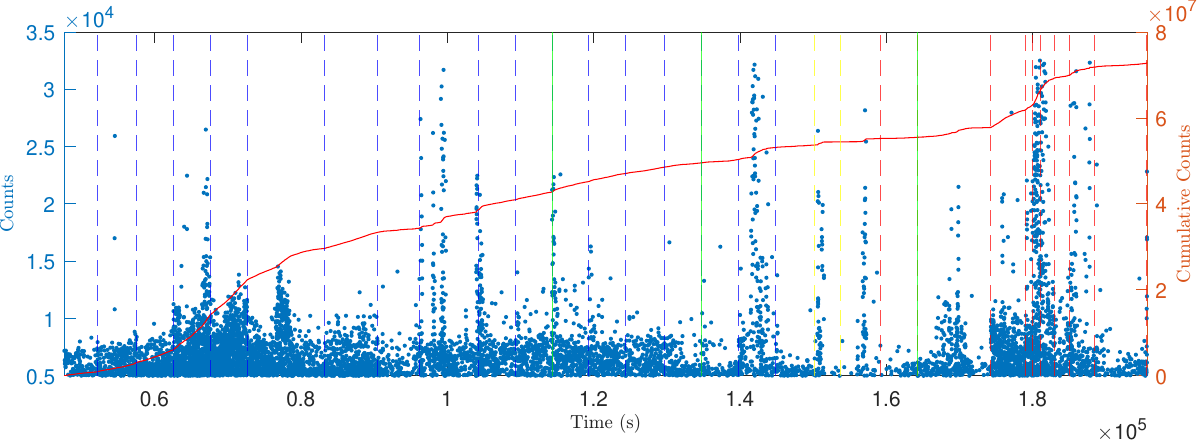} 
        \caption{}
        \label{fig:ReducedSet}
    \end{subfigure}
    \caption{Experimental results: (a) $10\%$ of complete dataset $(N=75,519)$ and (b) reduced dataset $(N=10,000)$. The retained data in the reduced set correspond to a subset starting at $\sim47,000s$ into the fatigue test.}
    \label{fig:fatigue_data}
\end{figure}

From the sensor-arrangement presented in Figure \ref{fig:sensor_arrangement}, only the waveforms recorded from Channel Five were used in this analysis, given that the corresponding sensor was the one placed nearest to the lug subjected to the fatigue load. For the duration of the fatigue test, this sensor accounted for a total of $755,193$ recorded waveforms. Handling this amount of data quickly became a complication, and it was necessary to extract features from the recorded waves in order to proceed with the analysis. Following the same reasoning as in the previous case study, the counts of each waveform were extracted and used as features for the analysis. In particular, a feature-vector $X=\{x_1,...,x_N\}$ representing the $N$ extracted counts was constructed for the development of the damage-detection algorithm. Figure~\mbox{\ref{fig:CompleteSet}} shows the evolution of the counts as the waveforms were recorded during the fatigue test. The dataset is rich in information, and it is not all that clear whether an obvious pattern exists that could indicate a form of fracture emerging and evolving until failure. Nevertheless, some peaks are observed more frequently towards the end of the test, suggesting that a more elaborate monitoring mechanism could anticipate failure and warn an operator of the urgency for an inspection of the affected part.

Using a collapsed Gibbs sampler to infer the cluster assignment has the advantage of evaluating each sample-point marginally. The implication of this advantage is that the model can adapt in real time as new sample-points are introduced, allowing for an online monitoring approach. However, as the dataset grows with the addition of new sample-points, inference becomes increasingly slower, to the point where inferring the posterior likelihood takes longer than the rate at which new observations are introduced. This drawback comes from having to reassess the assignment of every sample-point on each iteration of the Gibbs sampler. Because of this limitation, only a subset of the total number of recorded waveforms were considered. Indeed, an associated risk exists from potentially neglecting waveforms derived from structural damage, making this an important parameter to manage for this type of application. In this case, however, a reasonable balance between time of computation and quality of results was achieved by retaining $10,000$ samples (Figure \ref{fig:ReducedSet}).

An important step worth mentioning here is on dye-penetrant visual inspections that were carried out periodically throughout the experiment. The outcome of each visual inspection is represented by the colour of the dashed vertical lines included in the figures, where: (1) blue is for no evident fracture, (2) yellow is for possible fracture, and (3) red is for a confirmed surface fracture. Carrying out this simple step adds special value to this dataset, since it gives one means to validate the performance of a novelty detector. That is, the monitoring system should be expected to flag an abnormal operation prior to the first \say{possible fracture} given by the visual inspection. The green lines here indicate the instances when the applied load was increased.

In an ideal scenario, where background AE activity is almost non-existent, a sudden up-turn in energy should indicate anomalous activity, likely to have been caused by crack nucleation or growth. However, as demonstrated in this case, it is in fact difficult to discern any clear patterns in the dataset that can reliably indicate the presence of damage. The small load contribution administered by the sliding tube proves to have a substantial effect on the overall AE activity. Figure \ref{fig:CompleteSet} illustrates a clear example on how overwhelming AE datasets can get in practice.

The idea here is to have the DP-PMM cluster the waveforms by their counts. It is assumed that count-rates corresponding to anomalous AE events will somewhat deviate from those corresponding to benign events. By clustering the AE events, closer attention can be paid to the inferred clusters that are characterised by the growth and development of fracture. Other features from these clusters, such as energy, can then be monitored to determine the health-state of the lug.


\subsection{Damage-detection strategy}

There are different ways in which this method can be used for novelty detection. A simple mechanism could involve having the DP-based model trigger an alarm upon the formation of a new cluster. This approach would require having the model learn from a training set in which it is known that the structure is operating normally. During the training process, several clusters would form to account for the normal operation of the structure subjected to Environmental and/or operational Variations (EoVs), and any new clusters forming after the training phase would suggest an anomalous behaviour that has not yet been observed.

One immediate complication of this approach is that the formation of new clusters may not necessarily be a direct consequence of some change occurring to the physical system, and it may instead be attributed to the stochastic nature of the sampler when inferring the number of non-empty components~\cite{Rogers2016}. At each iteration, when the assignment of a sample-point is resampled, a finite probability of forming a new cluster exists, and the number of non-empty components may briefly grow as a result. These sporadic components may attract one or two members, but chances are that they will collapse again after a few iterations of the sampler. In practice, the new formation of clusters may only be an indicator of anomalous behaviour if the clusters \say{survive} to account for future observations. A possibly sensible alternative could be to instead monitor the rate at which the number of members from each component grows over time.

Cluster growth-rate can be a more robust indicator of damage, by working on the assumption that the generation of benign AE events will be steady. This assumption may hold since the cyclic load applied by the sliding tube remained unchanged for the duration of the experiment, and no additional inputs were introduced. Unless subjected to unforeseen variations, any clusters formed to characterise the benign AE events should therefore sustain a somewhat constant growth-rate. In contrast, a sudden release of energy caused by fracture will not only result in the additional formation of clusters, but also in a more inconsistent growth-rate of these, since AE events of this type will likely occur sporadically with the progression of the original fracture, or from the generation of new fractures in the structure.


\subsection{Results and discussion}

\begin{figure}[htbp]
    \centering
    
    \includegraphics[width=0.90\textwidth]{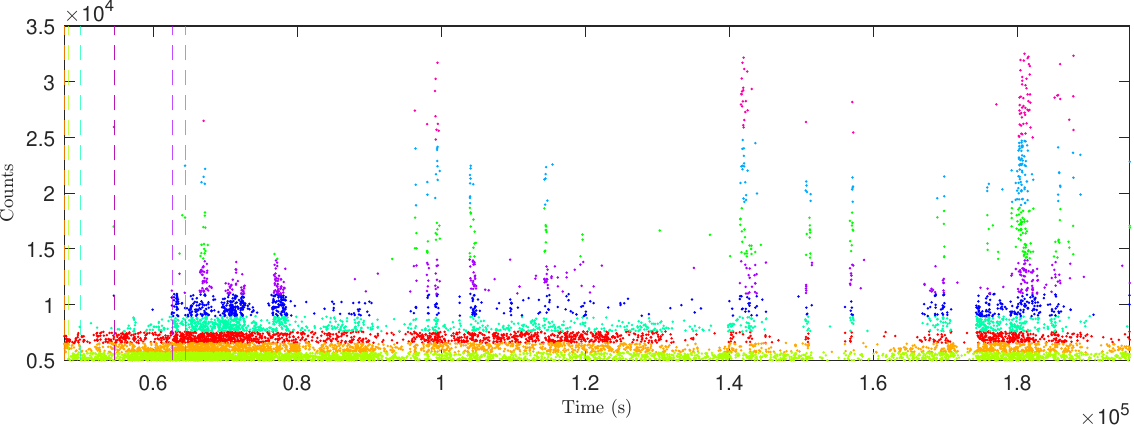} 
    \caption{AE counts clusters inferred by the DP-PMM.}
    \label{fig:p6_counts_clusters}
\end{figure}

Following the strategy described above, the DP-based model clustered the recorded AE waveforms as shown in Figure \ref{fig:p6_counts_clusters}. The dashed vertical lines indicate the moment when a new cluster was created. A different colour was used to represent each cluster and their corresponding members. The dataset was eventually split into $K = 9$ different non-empty components, by setting $\alpha=0.01$.

One of the first observations that can be made from the results presented in Figure \ref{fig:p6_counts_clusters}, is on the rate at which new clusters appeared. At the beginning of the fatigue test, the model was exposed (for the first time), to a vast number of AE events defined by a variety of different features. It is therefore natural to expect most clusters being created at the start, when the model is learning to identify and group the AE events as they are being introduced. Most of the first-appearing clusters are likely to correspond to the benign generating mechanisms attributed to the friction introduced by the sliding tube. Some events exhibiting significantly higher counts can also be observed occurring somewhat early in the test, also triggering the creation of new clusters.

These high-count events could potentially be attributed to sources of high energy, which may indicate the initial stages of crack initiation. Another plausible explanation is that these events might also be a result of internal fractures already existing prior to the start of the fatigue test. A point not yet mentioned, is that the lug had previously been fractured from preceding fatigue tests, and then repaired by welding it back in its original place. The sudden release of the internal stresses induced by the weld could have justifiably been a source of AE events exhibiting high counts. It is, unfortunately, impossible to know for sure the sources of all the observed events, as they could have also been generated from a variety of other mechanisms, such as material plastification, crack closure and crack-face rubbing~\cite{Roberts2003}, among others. 

Nonetheless, some consolation can be found by recalling that AE events deriving from crack nucleation and propagation will tend to exhibit a higher energy content (and counts), than those from any other potential mechanism in this setting. As reviewed in Section \ref{sec:intro}, the characterisation of AE events in fatigue tests is a subject that has been thoroughly examined in the literature, and on the basis of this consideration, one can therefore simplify the monitoring scheme by only looking at clusters that are characterised by the highest count-rates.

\begin{figure}[htbp]
    \centering
    
    \includegraphics[width=0.90\textwidth]{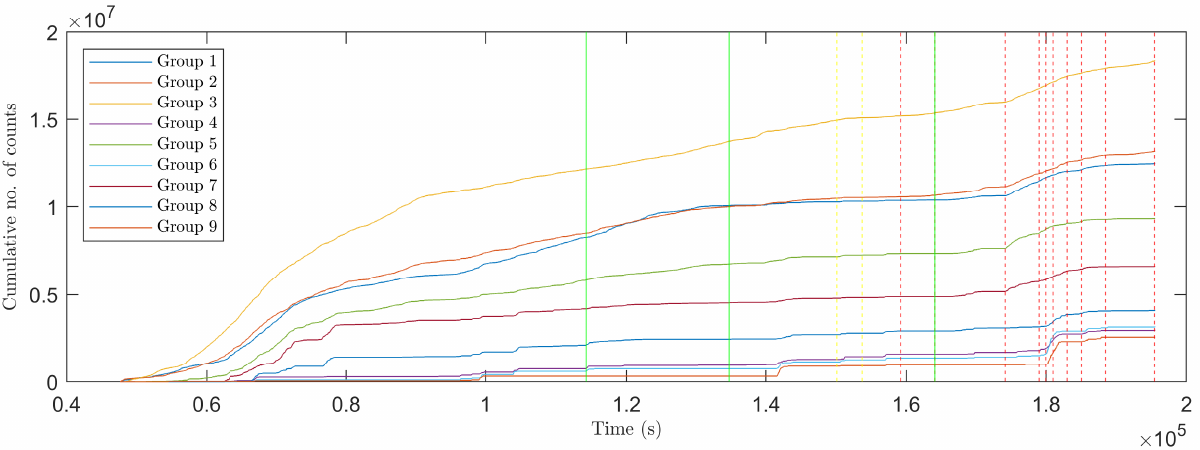} 
    \caption{Cumulative counts of all inferred groups.}
    \label{fig:CumulativeCounts}
\end{figure}

Figure \ref{fig:CumulativeCounts} shows the cumulative sum of events corresponding to each of the identified clusters. The counts of events were accumulated in this way to help visualise the growth-rate of the clusters throughout the test. A somewhat steady growth can be observed in almost all clusters, with a few exceptions exhibiting noticeable step increases. Since benign AE events are expected to emerge at a continuous rate, irrespective of the health-state of the lug, it may be reasonable to assume that clusters representing background activity are those that present a steady growth. Conversely, high-energy bursts manifesting from the growing crack are expected to be represented by the sudden step increases exhibited by some of these patterns.

\begin{figure}[htbp]
    \centering
    \includegraphics[width=0.90\textwidth]{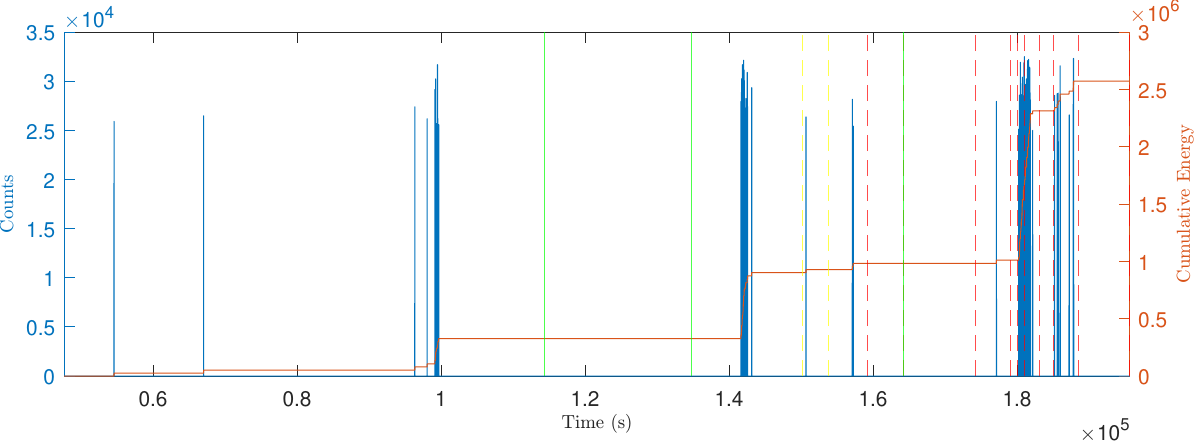} 
    \caption{Rate of counts corresponding to AE events assigned to Group $9$.}
    \label{fig:CountCumulativeSumGroup9}
\end{figure}

The clusters exhibiting these sharp up-turns were therefore extracted and inspected in isolation. The count rate of the $9^{th}$ cluster can be observed in Figure \ref{fig:CountCumulativeSumGroup9}. This cluster is in fact characterised by the AE events with the highest energy content, and its selection was based on this parameter. The results now clearly show peaks happening sporadically, with sharp up-turns manifesting throughout the duration of the experiment. By removing the background AE activity, the data become clearer and easier to interpret. Additionally, the size of the dataset is drastically reduced, and only the AE-waves of interest are retained, making it suitable for the development of a robust monitoring system.

Visually, one can see peaks manifesting moments before the visual inspection indicated a possible presence of fracture, where a first warning is provided in advance at $10,000s$, and the last by the prominent step increase occurring at approximately $140,000s$ into the experiment. The observed peak in this scenario may be interpreted as an indication of fracture extending to a detrimental length, and this form of warning is provided early enough for an operator to intervene and have the landing-gear inspected for damage.

The advantage of this approach is that the early presence of damage could be detected without needing to resort to source-localisation techniques, which can be cumbersome to implement, and require at least three sensors to be placed within the propagating medium of the generated waves. Here, only a single sensor was needed, and although the collected data from this sensor alone were very rich, the processing strategies presented in this work appear to be enough to have the model successfully flag an abnormal event in both real time and without having to perform an invasive procedure. Unfortunately, there is no way to characterise the sources of AE in this configuration, but the assumptions made here appear to satisfy the identification of AE-waves that were more sensitive to damage.


\section{Conclusions and Future Work}

The approach presented in this work demonstrated promising results in the probabilistic detection of AE events in time-series signals. Challenges associated with pre-defining an arbitrary hard-threshold were addressed. In the presented method, threshold-crossings in a given time window are no longer assumed to indicate, with absolute certainty, the existence of an AE event. Instead, the number of threshold-crossings given in a time window is treated as a random sample, which is then used to infer a probability of an event existing in that time window.

Given the probabilistic framework of the model, when an event is identified, a degree of (un)certainty on its assignment is provided. An operator could therefore decide to retain those AE events categorised with high certainty, after establishing a minimum criterion that is suitable for the application at hand.

Additionally, the proposed methodology gives some insight into scenarios where the nature of the AE activity is unknown, by not only detecting but also clustering all the observable events. Some form of validation would still be required to ensure that the threshold, window length, overlapped portion and hyperparameters are optimal. Nevertheless, it was demonstrated that these parameters can be marginalised out during the inference of the infinite mixture model. Challenges related to this aspect of the model were emphasised, particularly regarding the computational resources required for inference. To make the presented methods more practical and accessible, a more efficient approach to approximate the posterior would certainly be necessary. Addressing this issue is left as a subject for future research.

Finally, the promising capabilities of the DP-PMM in AE-based monitoring methods were explored further with experimental data gathered from a fatigue test of a landing gear. The implementation of the DP-PMM helped simplify the interpretation of an overwhelming dataset, allowing one to identify the groups of AE events that were more sensitive to damage, and therefore detect the early onset of fracture in the structure. Although this method yielded a successful outcome for this case study, exploring how well the DP-PMM generalises when faced in different applications is an exercise worth pursuing in the future.


\vspace{24pt}

\noindent \textbf{Acknowledgements} \vspace{12pt}

The authors gratefully acknowledge the support of the UK Engineering and Physical Sciences Research Council (EPSRC) via Grant references EP/R004900/1 and EP/W002140/1. The authors also gracefully acknowledge the incredible work of The University of Cardiff and Messier-Dowty Ltd.\ for their fatigue-testing design on the Airbus A320 main fitting, and for obtaining the provided AE data presented in this paper. For the purpose of open access, the authors have applied a Creative Commons Attribution (CC-BY-ND), licence to any Author Accepted Manuscript version arising.

\appendix
\section{Derivation of $p(z_{nk}=1|\boldsymbol{Z}^{-n},X^{-n},\boldsymbol{\alpha},a,b)$}
\label{appex:derivation_of_p_z}
Given the joint distribution defined by equation (\ref{eq:joint_post}), it is possible to derive the conditional distributions over $\boldsymbol{z}_n$, $\boldsymbol{\pi}$ and $\boldsymbol{\lambda}$. A conventional Gibbs sampler could then be employed to infer the joint distribution over these parameters. However, in this case, the probability distribution of interest is that defined over $\boldsymbol{z}_n$, conditioned on all other parameters. The motivation to derive an expression for $p(\boldsymbol{z}_n|...)$ begins by marginalising $\boldsymbol{\pi}$ and $\boldsymbol{\lambda}$ out of equation (\ref{eq:joint_post}). If $\boldsymbol{\pi}$ is first considered, then one can approach the marginalisation process in a two-step decomposition, where sampling from $p(\boldsymbol{z}_n,\boldsymbol{\pi})$ is said to be equivalent to drawing a sample of $\boldsymbol{\pi}$ first, and then using this sample to draw $\boldsymbol{z}_n$. That is,
\begin{align*}
    p(\boldsymbol{z}_n,\boldsymbol{\pi}|\boldsymbol{Z}^{-n},\boldsymbol{\lambda},\boldsymbol{\alpha},x_n) &\propto p(\boldsymbol{z}_n|\boldsymbol{\pi},x_n,\boldsymbol{\lambda})p(\boldsymbol{\pi}|\boldsymbol{Z}^{-n},\boldsymbol{\alpha}) \\
    &\propto {\displaystyle \prod_{k=1}^{K}} \left [\pi_k p(x_n|\lambda_k) \right ]^{z_{nk}} p(\boldsymbol{\pi}|\boldsymbol{Z}^{-n},\boldsymbol{\alpha})
\end{align*}
Here, $\boldsymbol{Z}^{-n}$ is the set of all assignments except the $n^{th}$ one. The notation of this expression can then be simplified by instead considering the value where the $k^{th}$ element of $z_n$ is equal to unity,
\begin{align*}
    p(z_{nk}=1,\boldsymbol{\pi}|\boldsymbol{Z}^{-n},\boldsymbol{\lambda},\boldsymbol{\alpha},x_n) &\propto \pi_k p(x_n|\lambda_k)p(\boldsymbol{\pi}|\boldsymbol{Z}^{-n},\boldsymbol{\alpha}) \\
    &\propto p(z_{nk}=1|\boldsymbol{\pi}) p(x_n|\lambda_k)p(\boldsymbol{\pi}|\boldsymbol{Z}^{-n},\boldsymbol{\alpha})
\end{align*}
In this form, it becomes possible to marginalise $\boldsymbol{\pi}$ out,
\begin{align*}
    p(z_{nk}=1|\boldsymbol{Z}^{-n},\boldsymbol{\lambda},\boldsymbol{\alpha},x_n) &\propto \int \! p(z_{nk}=1|\boldsymbol{\pi}) p(x_n|\lambda_k) p(\boldsymbol{\pi}|\boldsymbol{Z}^{-n},\boldsymbol{\alpha}) \, \mathrm{d}\boldsymbol{\pi} \\
    &\propto p(x_n|\lambda_k) \int \! p(z_{nk}=1|\boldsymbol{\pi}) p(\boldsymbol{\pi}|\boldsymbol{Z}^{-n},\boldsymbol{\alpha}) \, \mathrm{d}\boldsymbol{\pi} \\
    &\propto p(x_n|\lambda_k) p(z_{nk}=1|\boldsymbol{Z}^{-n},\boldsymbol{\alpha})
\end{align*}
where,
\begin{align*}
    p(z_{nk}=1|\boldsymbol{Z}^{-n},\boldsymbol{\alpha}) &= \int \! p(z_{nk}=1|\boldsymbol{\pi}) p(\boldsymbol{\pi}|\boldsymbol{Z}^{-n},\boldsymbol{\alpha}) \, \mathrm{d}\boldsymbol{\pi}
\end{align*}
and the posterior, $p(\pi|\boldsymbol{Z}^{-n},\boldsymbol{\alpha})$, is the result of having a Dirichlet prior, $p(\boldsymbol{\pi}|\boldsymbol{\alpha})$, as the conjugate of the multinomial likelihood, $p(\boldsymbol{Z}^{-n}|\boldsymbol{\pi})$. Concretely,
\begin{align*}
    p(\boldsymbol{\pi}|\boldsymbol{Z}^{-n},\boldsymbol{\alpha}) \propto {\displaystyle \prod_{k=1}^{K}} \pi_k^{\beta_k-1}, \quad \beta_k = \alpha_k + c_k^{-n}
\end{align*}
where $c_k^{-n}$ is the number of objects assigned to the $k^{th}$ group, excluding the assignment on $x_n$. Hence,
\begin{align*}
    p(z_{nk}=1|\boldsymbol{Z}^{-n},\boldsymbol{\alpha}) &= \frac{\Gamma (\sum_{j=1}^K\beta_j)}{\prod_{j=1}^K\Gamma(\beta_j)} \int \! \pi_k \prod_{j=1}^{K} \pi_j^{\beta_j-1} \, \mathrm{d}\boldsymbol{\pi} \\
    &= \frac{\Gamma(\sum_{j=1}^K\beta_j)}{\prod_{j=1}^K\Gamma(\beta_j)} \int \! \prod_{j=1}^{K} \pi_j^{\beta_j+\delta_{jk}-1} \, \mathrm{d}\boldsymbol{\pi} \\
    &= \frac{\Gamma(\sum_{j=1}^K\beta_j)}{\prod_{j=1}^K\Gamma(\beta_j)} \frac{\prod_{j=1}^K\Gamma(\beta_j + \delta_{jk})}{\Gamma(\sum_{j=1}^K(\beta_j + \delta_{jk}))}
\end{align*}
where $\delta_{jk} = 1$ when $j=k$, and zero otherwise. Since $\sum_{j=1}^K\delta_{jk}=1$,
\begin{align*}
    p(z_{nk}=1|\boldsymbol{Z}^{-n},\boldsymbol{\alpha}) &= \frac{\Gamma(\sum_{j=1}^K\beta_j)}{\prod_{j=1}^K\Gamma(\beta_j)} \frac{\prod_{j=1}^K\Gamma(\beta_j + \delta_{jk})}{\Gamma(\sum_{j=1}^K\beta_j + 1)} \\
    &= \frac{\beta_k}{\sum_{j=1}^K\beta_j}
\end{align*}
Now, because $\beta_k=c_k^{-n} + \alpha_k$,
\begin{align*}
    p(z_{nk}=1|\boldsymbol{Z}^{-n},\boldsymbol{\alpha}) &= \frac{c_k^{-n} + \alpha_k}{\sum_{j=1}^K c_j^{-n} + \alpha_j}
\end{align*}
Following the same reasoning, it is also possible to collapse $\lambda_k$ from the sampler. The joint density over $z_{nk}=1$ and $\lambda_k$ is derived by carrying out an equivalent two-step decomposition, 
\begin{align*}
    p(z_{nk}=1,\lambda_k|\boldsymbol{Z}^{-n},X^{-n},\boldsymbol{\alpha},a,b) &\propto p(z_{nk}=1|\boldsymbol{Z}^{-n},\boldsymbol{\alpha})p(x_n|\lambda_k)p(\lambda_k|\boldsymbol{Z}^{-n},X^{-n},a,b)
\end{align*}
where $X^{-n}$ corresponds to all objects except for $x_n$. Marginalising $\lambda_k$ out from this expression gives,
\begin{align*}
    p(z_{nk}=1|\boldsymbol{Z}^{-n},X^{-n},\boldsymbol{\alpha},a,b) &\propto \int \! p(z_{nk}=1|\boldsymbol{Z}^{-n},\boldsymbol{\alpha})p(x_n|\lambda_k)p(\lambda_k|\boldsymbol{Z}^{-n},X^{-n},a,b) \, \mathrm{d}\lambda_k\\
    &\propto p(z_{nk}=1|\boldsymbol{Z}^{-n},\boldsymbol{\alpha}) \int \! p(x_n|\lambda_k)p(\lambda_k|\boldsymbol{Z}^{-n},X^{-n},a,b) \, \mathrm{d}\lambda_k\\
    &\propto p(z_{nk}=1|\boldsymbol{Z}^{-n},\boldsymbol{\alpha}) p(x_n|\boldsymbol{Z}^{-n},X^{-n},a,b)
\end{align*}
where,
\begin{align*}
    p(x_n|\boldsymbol{Z}^{-n},X^{-n},a,b) = \int \! p(x_n|\lambda_k)p(\lambda_k|\boldsymbol{Z}^{-n},X^{-n},a,b) \, \mathrm{d}\lambda_k
\end{align*}
and, like in the previous case, the posterior, $p(\lambda|\boldsymbol{Z}^{-n},X^{-n},a,b)$, is the result of having a Gamma prior $p(\lambda_k|a,b)$, as conjugate of the Poisson likelihood, $p(X^{-n}|\lambda_k,\boldsymbol{Z}^{-n})$. Concretely,
\begin{align*}
    p(\lambda_k|\boldsymbol{Z}^{-n},X^{-n},a,b) &\propto \lambda_k^{\delta-1} e^{-\gamma\lambda_k},
\end{align*}
\begin{align*}
    \delta = a + \sum_{m \ne n}z_{mk}x_m, \quad \gamma = \sum_{m \ne n}z_{mk} + b
\end{align*}
Hence,
\begin{align*}
    p(x_n|\boldsymbol{Z}^{-n},X^{-n},a,b) &= \frac{\gamma^\delta}{x_n!\Gamma(\delta)} \int \! \lambda_k^{(x_n+\delta)-1} e^{-(\gamma+1)\lambda_k} \, \mathrm{d}\lambda_k \\
    &= \frac{\gamma^\delta}{x_n!\Gamma(\delta)} \frac{\Gamma(x_n+\delta)}{(1+\gamma)^{(x_n+\delta)}}
\end{align*}
Re-arranging and having $r=\delta$ and $p=\gamma/(\gamma+1)$,
\begin{align*}
    p(x_n|\boldsymbol{Z}^{-n},X^{-n},a,b) &= \frac{\Gamma(x_n+r)}{x_n!\Gamma(r)} \frac{\left ( \frac{p}{1-p} \right )^{r}}{\left ( 1+\frac{p}{1-p} \right )^{(x_n+r)}}\\
    &= \frac{\Gamma(x_n+r)}{x_n!\Gamma(r)}p^r(1-p)^r \\
    &= NB(r,p)
\end{align*}
Finally,
\begin{align*}
    p(z_{nk}=1|\boldsymbol{Z}^{-n},X^{-n},\boldsymbol{\alpha},a,b) \propto \frac{c_k^{-n} + \alpha_k}{\sum_j^K c_j^{-n} + \alpha_j}p(x_n|\boldsymbol{Z}^{-n},X^{-n},a,b)
\end{align*}

\newpage
\section{Marginalisation of the threshold}
\label{appex:thresh_marginalisation}

\begin{figure}[htbp]
    \centering
    \includegraphics[width=0.9\textwidth]{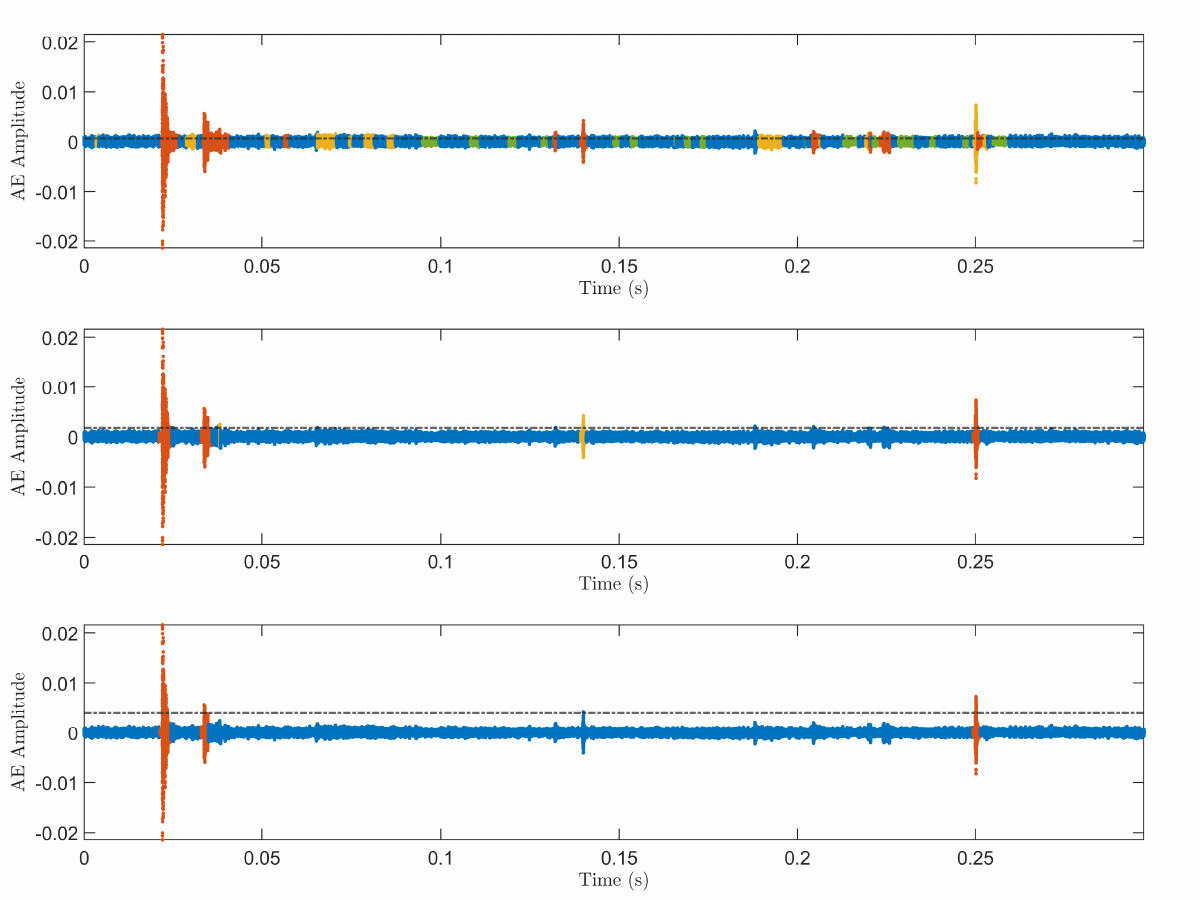}

    \caption{Assignment of sections in the signal to the groups that maximise the assignment probabilities, given thresholds corresponding the (Top) $95^{th}$, (Middle) $99.5^{th}$, and (Bottom) $99.9^{th}$ percentiles, respectively.}
    \label{fig:thresh123}
\end{figure}

\begin{figure}[htbp]
    \centering
    \includegraphics[width=0.9\textwidth]{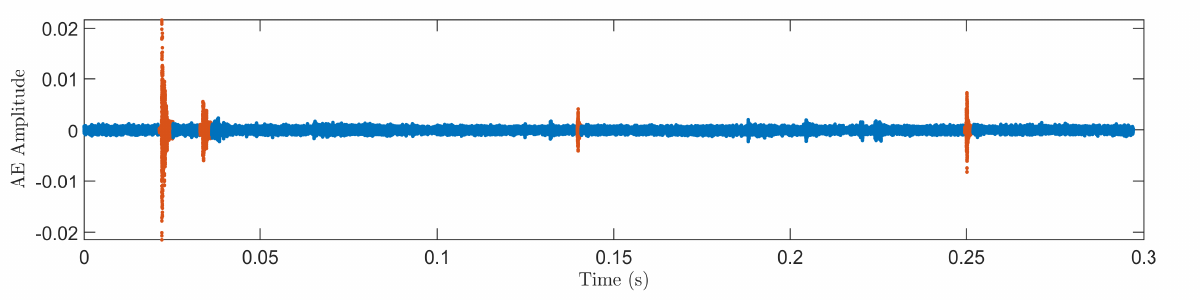}

    \caption{Assignment of sections in the signal to the groups that maximise the assignment probabilities after marginalising the threshold. From all the clusters inferred over each model, only two distinct groups of AE activity are retained from the marginalisation.}
    \label{fig:threshAvg}
\end{figure}

A salient aspect of Bayesian modelling is the ability to compare models in a principled manner. In this case, the threshold can be interpreted as the realisation of a model $\mathcal{M}_j$. By following the same approach as in Section \ref{sec:results_and_discussion}, the threshold can be marginalised out of the DP-PMM, thereby improving its overall robustness. In this demonstration, three sets of features are considered, extracted using three different thresholds. The thresholds used correspond to the $95^{th}$, $99.5^{th}$, and $99.9^{th}$ percentiles of the signal, respectively. For each model, the overlapping scheme from Section \ref{sec:results_and_discussion} is applied, resulting in a total of $24$ sets of features. The assigned sections of the signal, for each threshold, are shown in Figure \ref{fig:thresh123}. The results clearly demonstrate that a lower threshold leads the DP-PMM to cluster sections in the background noise, while a higher threshold disregards potential events in the signal. By averaging the probabilities obtained from each model, as described in Equation (\ref{eq:marginalisation}), the discrepancies across models are mitigated, and returns a more balanced outcome that is less sensitive to the chosen threshold. Figure \ref{fig:threshAvg} illustrates this outcome, where two distinct groups appear to dominate across all the clusters inferred from each model.

\newpage

\renewcommand{\bibname}{References}

\bibliographystyle{unsrt}

\bibliography{refs}

\end{document}